\documentclass[%reprint,
superscriptaddress,
 amsmath,amssymb,
 aps,
]{revtex4-1}

\usepackage{graphicx}% Include figure files
\usepackage{natbib}
\usepackage{upgreek}
\usepackage{dcolumn}% Align table columns on decimal point
\usepackage{bm}% bold math
\begin{document}

\preprint{APS/123-QED}

\title{Pair Interaction of Catalytically Active Colloids: From Assembly to Escape}% Force line breaks with \\

\author{Nima Sharifi-Mood}
 \affiliation{Chemical and Biomolecular Engineering, University of Pennsylvania, Philadelphia, PA, 19104, USA}
 \email{nimash@seas.upenn.edu}
\author{Ali Mozaffari}%
\affiliation{Benjamin Levich Institute and Department of Chemical Engineering, City College of the City University of New York, New York, NY 10031, USA}%
\author{Ubaldo M. C\'ordova-Figueroa}
\affiliation{Department of Chemical Engineering, University of Puerto Rico -- Mayag\"uez, Mayag\"uez, PR 00681, Puerto Rico}%
\begin{abstract}
The dynamics and pair trajectory of two self-propelled colloids are reported. The autonomous motions of the colloids are due to a catalytic chemical reaction taking place asymmetrically on their surfaces that generates a concentration gradient of interactive solutes around the particles and actuate particle propulsion. We consider two spherical particles with symmetric catalytic caps extending over the local polar angles $\theta^1_{cap}$ and $\theta^2_{cap}$ from the centers of active sectors in an otherwise quiescent fluid. A combined analytical-numerical technique was developed to solve the coupled mass transfer equation and the hydrodynamics in the Stokes flow regime. The ensuing pair trajectory of the colloids is controlled by the reacting coverages $\theta^j_{cap}$ and their initial relative orientation with respect to each other. Our analysis indicates two possible scenarios for pair trajectories of catalytic self-propelled particles: either the particles approach, come into contact and assemble or they interact and move away from each other (escape). For arbitrary motions of the colloids, it is found that the direction of particle rotations is the key factor in determining the escape or assembly scenario. Based on the analysis, a phase diagram is sketched for the pair trajectory of the catalytically active particles as a function of active coverages and their initial relative orientations. We believe this study has important implications in elucidation of collective behaviors of auotophoretically self-propelled colloids. 
\end{abstract}
\maketitle
\section{Introduction}
Directed motion of colloidal particles via self-created gradients or fields has become an area of significant scientific interest and potential. Drug or gene delivery \cite{Coti_Drug,Wang_Drug,Wang_Magnet}, neutralizing pollutants \cite{Wang_Pul1,Wang_Pul2}, directed transport of chemical reagents \cite{Sen_Cargo,Wang-Cargo}, and mixing enhancement \cite{Wang_mixing} can be mentioned here. With recent developments in the collective behavior of self-propelling particles \cite{Bocquet,Palacci_Pine,Speck}, the interest has shifted towards creating active soft materials that, in addition to operating autonomously, also have the ability to sense their surroundings and process this information into smart decisions.

After a decade, it is now well understood that an asymmetric surface chemical reaction on a colloidal particle is able to produce autonomous motion. This observation has been studied both experimentally \cite{Paxton2004,Paxton2006,Howse2007} and theoretically \cite{PhysRevE.81.065302,posner2011,Golestanian2007,Cordova-Figueroa}. In the particular case of non-electrolyte solutions, Golestanian \emph{et~al.} \cite{Golestanian2007,Howse2007} utilized a continuum description based on phoresis theory \cite{AP84,a89} to study the locomotion of spherical Janus particles due to a surface chemical reaction. In this case, a chemical reaction takes place at a portion of the particle surface and produces a nonuniform distribution of the product and reactant solutes which in turn induces a diffusiophoretic motion \cite{AP84}. A similar idea was applied for a thermophoretically self-propelled Janus particle, which consumes an imposed laser energy on half of its surface, thus creating a nonuniform temperature field \cite{Jiang}.  In contrast, C\'{o}rdova-Figueroa and Brady \cite{Cordova-Figueroa,Brady,Ubaldo_Flux} applied a colloidal perspective to the problem of catalytically driven self-propulsion, in which a catalytic motor (or motors) does not have to be assumed to be large in comparison with the product (or reactant) solutes and the restriction for diluteness of solutes can be relaxed. Therefore, the results obtained within this approach can be applied to other microscopic objects, such as proteins or even large molecules.  

Current efforts in the study of catalytic motors include exploiting propulsion mechanisms and finding alternatives to guide their direction \cite{ssss2011,Wall_1}.  Moreover, self-propulsion was found even for a symmetric chemical reaction  \cite{Popescu1,Bartolo} within a chemotaxis problem \cite{Grima,Sengupta}. There, self-propulsion emerges as a result of the instability of a stationary state; its direction is determined by an initial perturbation in the solute concentration gradient.  For example, the presence of a chemically passive particle (or cargo) connected to a catalytic one causes self-propulsion \cite{Kapral1,Popescu1,Nima_thesis,Kapral2}. This model provides a good representation of a cargo particle transported by the catalytic motor, a problem of crucial importance for a large number of applications. On a similar note, Shklyaev \emph{et~al.} \cite{Chemical_Sailing} and Michelin and Lauga \cite{Michelin2} reported that self-propulsion of a particle with a uniform distribution of a surface reaction arises due to nonsphericity of the particle shape.  A direct significance of this study is the realization that even for particle with uniform surface catalytic activity, a change to the surrounding distribution of reagents (or solutes) can occur which, consequently, creates perturbations (or concentration gradients) that directly affect particle diffusivity and the interaction force between nearby particles.

An increasing number of experiments on catalytically-driven (active) colloidal particles have shown that the interaction of chemically active particles is more complicated than usual interaction of two nonreactive (passive) particles. Indeed, each chemically active particle changes the distribution of product (or reactant) solutes which, in turn, alters the motion of the other particles. In this case, the motions of active particles are influenced not only by hydrodynamic interactions but also via diffusive interactions of the solute distribution generated (or consumed) at the surface of the active areas of each colloid with the boundary of other colloids. Such is the case in the problem of multicomponent diffusion where a gradient in the concentration of one species can drive the flux of another \cite{Bird,Batchelor1}. Similarly, depletion flocculation occurs because small particles (e.g., polymers) are excluded from a zone separating two nearly touching colloidal particles and the imbalanced osmotic pressure of the small particles causes an entropic attractive force leading to flocculation  \cite{Asakura,Jenkins}. Depleted regions between two or more catalytically active colloids can unite and hinder or stop their autonomous motion. 

Also, it has been reported recently that active suspensions exhibit perplexing collective behaviour such as swarming, predator-prey interactions and chemotaxis which are reminiscent of biological microorganisms  \cite{Velegol,Ibele,Aranson,Lowen}. Active catalytic microswimmers have been experimentally observed to form clusters \cite{Velegol,Bocquet} which are ``live'', i.e. the particles join and leave the clusters frequently. Moreover, the experimental evidence indicates in addition to the large clusters, there are coexistent gas phases where the particles swim freely. In dilute suspensions, it was shown that the mean cluster size increases linearly with propulsion speed \cite{Speck}. This behavior has been observed for both purely repulsive and attractive suspensions with the only difference being the mean cluster size (larger for attractive particles).  This clustering and phase separation have also been investigated via simulations \cite{Speck,Baskaran}. The clustering and phase separation in active matter is driven by non-equilibrium rather than attractive interparticle forces. Cates and Tailleur \cite{Cates3} have theoretically examine a phase separation for a model based on a run-and-tumble motion. They showed that a locally reduced mobility can result in dense regions where the directed motions of swimmers are blocked by their neighbors and a dilute gas of autonomously moving particles.\\
A  ``micromotor'' move in random directions at speeds of tens of microns per second. But collectively, catalytically-driven colloids immersed in a Newtonian fluid are expected to behave as a complex fluid, but not in a traditional fashion  \cite{Aranson,Yan-Brady}. The interaction of catalytically-driven colloidal particles is expected to be largely mediated by the intervening fluid, hence, insight into the hydrodynamics associated with the motion of two particles is necessary for understanding their collective behavior. This is apparent in a recent study in which it was show that the near field hydrodynamic interaction acting between squirmers and confining walls is crucial to determine their collective motion  \cite{Stark}. Additionally, unlike a suspension of squirmers, the collective behavior of a suspension of catalytic particles can also be drastically influenced by phoretic effects \cite{Palacci_Pine}, and hence a pure hydrodynamic model (such as a squirmer model) may not be adequate to capture the collective dynamics.

In this study, we take the first step to understand the collective behavior and dynamics of a suspension of catalytically active self-propelled particles. In so doing, we probe the pair interaction of two catalytically active colloids from a continuum framework in the context of self-diffusiophoresis \cite{a89,Howse2007}. This mechanism relies on an unbalanced intermolecular interaction in a thin interaction layer $L$ caused by an asymmetric surface reaction around the colloid and thus results in a propulsion. In the limit of small interaction layer compared to the particle size (i.e. $L\ll a$), it has been shown \cite{a89} that the intermolecular interaction generates a slip velocity ${\bf{v}}_{slip}$ on the particle surface which scales with the local solute concentration, ${\bf{v}}_{slip}=-b\nabla_s n_s$, where $b$ is the mobility coefficient, $\nabla_s$ is the surface gradient operator and $n_s$ is the solute concentration evaluated at the particle surface (see below for definitions). The validity of this approach has been examined from a micromechanical perspective \cite{Brady} and with molecular dynamics simulations \cite{Nima_2}. Thus, to capture the ``swimming'' (self-diffusiophoretic) velocity of the particle, the equations governing solute distribution (microstructure) and hydrodynamics must be solved simultaneously. The self-diffusiophoretic velocity of a non-Brownian catalytically active colloid in an infinite medium scales with the Damk\"{o}hler number, $Da=ka/D_{A}$, and the P\'eclet number, $Pe=U_{c}a/{D_A}$, where the former determines the relative importance of reaction rate to the solute diffusion and the later describes the ratio of the solute advection to the solute diffusion and $k$ is the first order kinetic constant, $D_{A}$ is the reactant solute diffusivity, $a$ is the particle radius and $U_c$ is the characteristic swimming velocity of the particle. For the reactivity of the particles, a simple surface reaction model $A\to B$ is assumed, where $A$ and $B$ represent the reactant and product solutes, respectively, taking place uniformly solely on the cap region $0<\theta\le \theta_{cap}$ of the colloid and brings about a constant flux production of the solute $B$, $N_0$, on the reactive cap and elsewhere, there is a zero flux condition since the colloid is impenetrable to the solutes. The constant flux production assumption has been shown to be legitimate when considering a surface reaction with first order kinetics and $Da\ll 1$ (see \cite{Nima_1,Ubaldo_Flux,Lauga_Peclet} for details). Additionally, at $Pe\ll 1$, the hydrodynamics and the equation governing the solute distribution are decoupled. Utilizing the slip velocity argument, the solute distribution can be obtained by solving a Laplace equation subjected to a relevant boundary condition on the colloid surface and thereafter hydrodynamics can be solved to yield the dimensional self-diffusiophoretic velocity as \cite{Howse2007},
\begin{eqnarray}
{U_\infty } = b\frac{{{N_0}}}{{{D_A}}}\left[ {\frac{{1 - {{\cos }^2}{\theta _{cap}}}}{4}} \right].
\end{eqnarray}
Note that the above relation is deterministic, i.e. the time scale for Brownian rotation of the particle $1/D_r$ ($D_r$ is the rotational diffusivity of the particle) has been considered to be much longer than the time scale for directed motion, $a/U_{\infty}$. 

This continuum approach has been used in a number of studies to probe the diffusiophoretic autonomous motion of a spheroidal particle \cite{refId}, a spherical particle near a planar solid wall \cite{Uspal,Wall_1} and in a simple shear flow \cite{Frankel}. Furthermore, the effect of finite rates of solute advection and surface reaction on the propulsion speed have been addressed in the literature \cite{golestanian_size,Ubaldo_Flux,Nima_thesis,Lauga_Peclet}.

Here, a similar approach is used to address the pair interaction of two partially active colloids with arbitrary orientations. The orientation of each colloid is defined via a unit vector directed from the active to the passive poles of the colloid.  Continuum calculations are undertaken in which the repulsive interactions of the solute product with the colloids actuate the autonomous motions, and the reactive faces produce a constant flux of product solutes. The solute distribution around the particles is found by solving the diffusion equation exactly in a curvilinear bispherical coordinate. The translational and angular velocities of the particles at the Stokes flow regime are obtained using the Reynolds Reciprocal Theorem (RRT)  \cite{hapbren83} based on an asymptotic approach in which the net interaction creates a slip velocity at the surface that actuates the motion \cite{a89}. We consider first the axisymmetric locomotion of two partially active colloids along their line of centers where the active face of one colloid is oriented either directly in front of, or facing away from, the active (or passive) section of the other colloid. In this case, depending on the pair orientations, the colloids either move toward or away from each other without rotation. On the other hand, for non-axisymmetric motion, the analysis indicates that when two catalytically active colloids approach, they undergo rotations around axes perpendicular to the line of centers. The magnitude and direction of rotations are functions of the colloids separation distance, relative orientation and the areas of active sections, i.e. the colloids can rotate in such a way that the active surfaces face towards or away from each other and, consequently, the colloids assemble or escape. We discuss the hydrodynamics of diffusiophoretic motion of two colloids and demonstrate under what circumstances the active colloids can assemble or escape. This study is crucially important to illuminate the role of hydrodynamic interactions on collective behaviors of the self-propelled particles as these interactions are often neglected in theoretical studies and numerical simulations \cite{Palacci_Pine,Speck}. Moreover, the current study reveals a unique feature of suspensions of diffusiophoretically self-propelled particles compared to other active systems, namely ``chemical signaling'', i.e. how the chemically active particles can be pushed (or pulled) away from regions where the local concentration of solute is high (or low) due to diffusive nature of the solute. 

The paper is organized as follows: In \S~\ref{S1}, we provide the formulation and the solution technique developed to solve the mass transfer and hydrodynamic equations and in \S~\ref{S2} the results and a discussion of the concentration field, particle velocities and pair trajectories of the catalytically driven colloids are given. Concluding remarks are provided in \S~\ref{Conc}.

 \begin{figure} 
\centering
\includegraphics[width=0.5 \textwidth]{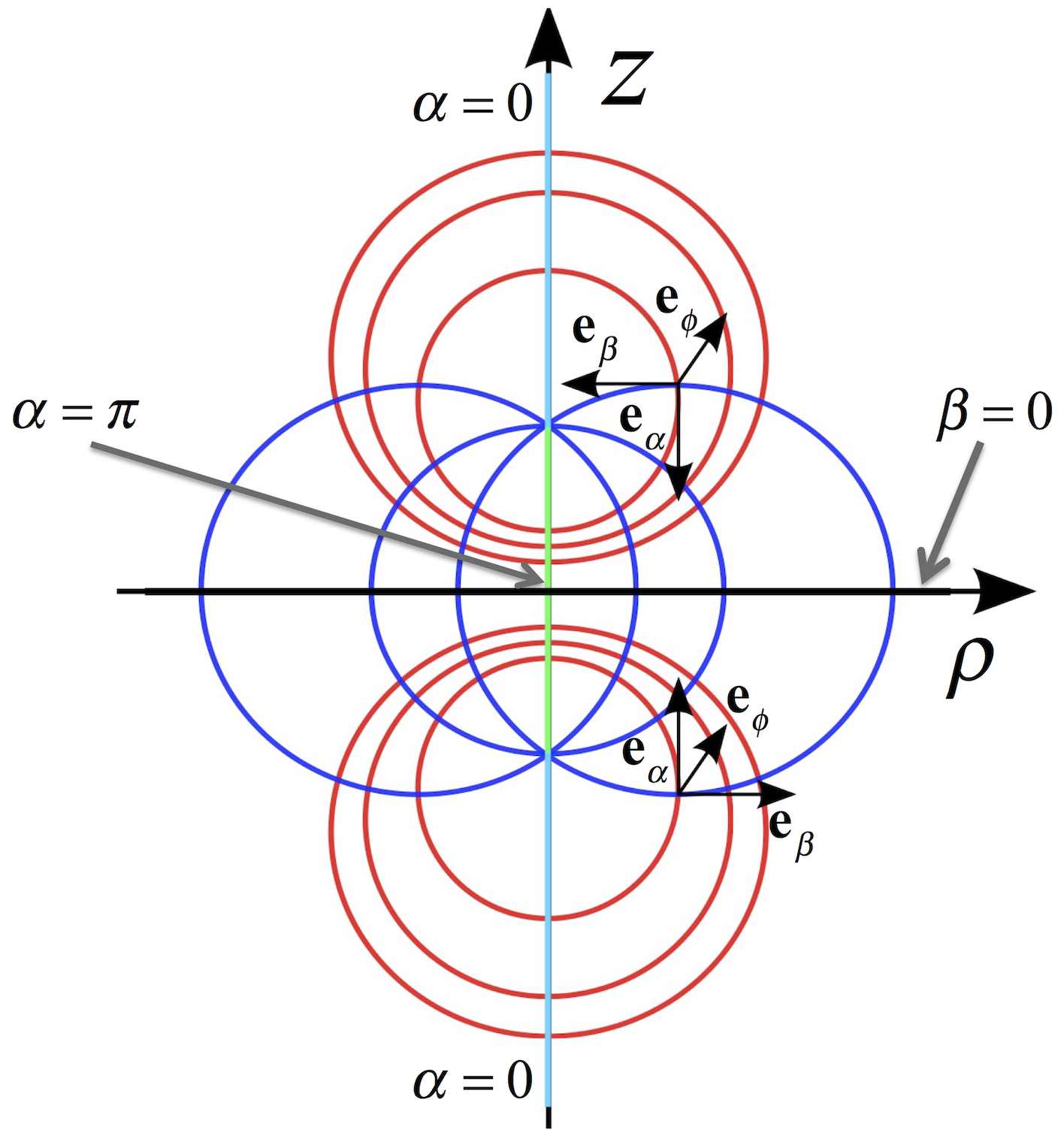}	     	    
 \caption{\footnotesize{ Geometrical sketch of bispherical coordinates $(\alpha ,\beta ,\phi)$.} } 
\label{bis2}
 \end{figure} 

\section{Formulation and analytical solution}\label{S1}
\begin{figure} 
\centering
\includegraphics[width=0.35 \textwidth]{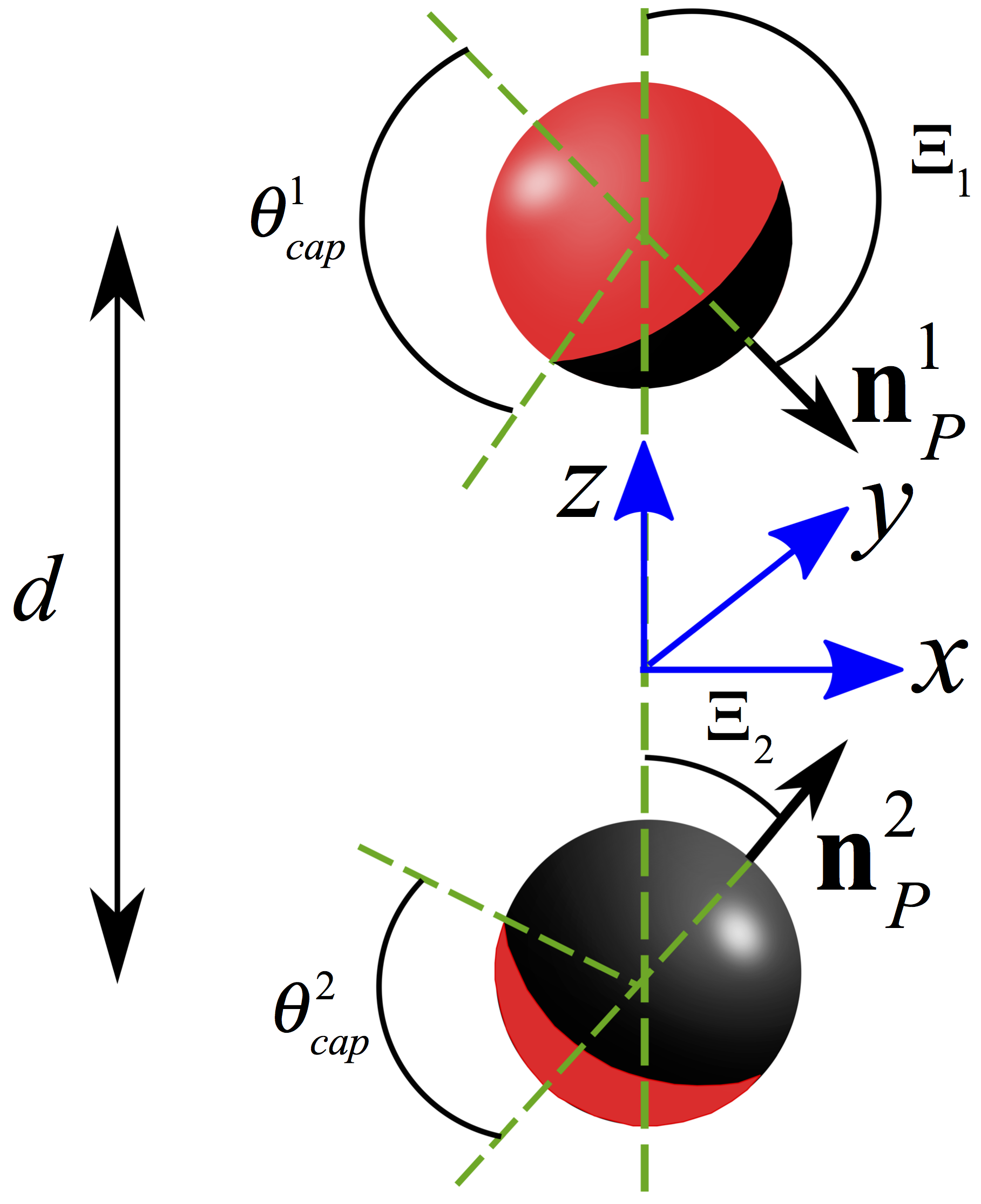}	     	    
 \caption{\footnotesize{Schematic representation of two catalytically active motors where $\theta_{cap}^j$ accounts for the extent of the active region for the colloids, ${\bf{n}}_{P}^{j}$ is the unit normal directing from active to passive poles and $\Xi_{j}$ determines the relative orientation of the two colloids ($j=1$ and $2$).} } 
\label{Fig_1}
 \end{figure}

Since the key feature here for autonomous motion is the concentration field around the colloids, the solute distribution in the presence of both colloids must be captured. Thus, the mass transfer equation, which is a Laplace equation can be solved in a set of orthogonal curvilinear bispherical coordinates $(\alpha ,\beta ,\phi)$ (see Fig.~\ref{bis2}). We assume the colloids have similar radii $a$ and their center-to-center distance is denoted by $d=a\Delta$. The colloids are partially active and the extent of their active regions is defined by two angles, $\theta_{cap}^{1}$ and $\theta_{cap}^{2}$, which vary between $0^{\circ}$ and $180^{\circ}$, where $0^{\circ}$ accounts for a passive particle while $180^{\circ}$ indicates a colloid that is catalytically active over its entire surface. Furthermore, to investigate the relative orientation of the particles, we define a unit orientation vector associate with each colloid directing from the active to the passive poles. The inclination angles of the two colloids, $\Xi_1$ and $\Xi_2$, can be considered as angles between their unit orientation vectors and the positive direction of center-to-center line ($z$-direction); these angles varies between $0^{\circ}$ and $360^{\circ}$ (see Fig.~\ref{Fig_1}). The motion of the particles is considered to be non-Brownian ($t \ll  1/D_r$) and pseudosteady state and the hydrodynamics is in the Stokes flow regime where the inertial terms can be neglected.
\subsection{Concentration field of product solute}
A simple form for the activity of the colloidal particles is taken which consists of a constant flux production of solutes uniformly distributed on the catalytic surface. This assumption has been shown to be valid in the limits of small rate of reaction compared to the rate of solute diffusion, i.e. $Da\ll 1$. For small solute in infinite dilution limit, the non-dimensional steady concentration field of the solute product at $Pe \ll 1$ can be written as,
\begin{eqnarray}
{\nabla ^2}n = 0,\label{Laplacian}
\end{eqnarray}
with the following boundary conditions,
\begin{eqnarray}
&{\bf{e}}_n^j\cdot\nabla n =  - {\vartheta _j}u_j({\bf{r}}),~~~{\bf{r}} \in {S_c^j},~(j = 1,2),\\
&n \to 0,~~~~~~~(\left| {\bf{r}} \right| \to \infty ),
\end{eqnarray}
where ${\bf{e}}_n^j$ is the unit normal vector pointing outward to the surface of the $j$th colloid, $n$ is the non-dimensional concentration of the product solute, ${\bf{r}}$ is the non-dimensional position vector with respect to the $j$th colloid center, $S_c^j$ accounts for the surface domain of the  $j$th colloid, $u_j$ is the coverage function of colloid $j$ which is one in its active section and zero elsewhere and ${\vartheta _j}$ represents the scaled solute flux production for colloid $j$ defined as,
\begin{eqnarray}
{\vartheta _j} = \frac{{{N_j}}}{{{N_1}{}}},
\end{eqnarray}
where $N_j$ denotes the dimensional solute flux production for colloid $j$. All variables mentioned above were non-dimensionalized according to the characteristic concentration of the solute $n_c\sim {{{N_1}{a}}}/{{{D_A}}}$ and particle radius $a$. To satisfy the boundary conditions at both colloid surfaces, we utilize a bispherical coordinate $(\alpha ,\beta ,\phi)$. Bispherical coordinates are related to cylindrical coordinates ($\rho,\phi,z$) with the following transformation
\begin{eqnarray}
\displaystyle{\rho = \frac{{c \;\sin \alpha \;}}{{\cosh \beta  - \cos \alpha }}},\label{p14}\\
\displaystyle{z = \frac{{c \;\sinh \beta }}{{\cosh \beta  - \cos \alpha }}}.\label{p15}
\end{eqnarray} 
Note that the $z$-axis is always aligned in the direction of the center-to-center line of the particles. The range of coordinates are limited to $0 \leqslant \alpha  \leqslant \pi $, $\beta_2< \beta < \beta_1 $ and $0\leqslant \phi  \leqslant2\pi $, for positive scale factor $c$. Furthermore, $\beta  = \beta _1>0 $ represents a sphere whose center is located on the positive $z$-axis at point $z = c~\coth \beta _1 $ with radius of $r_{c1}=c/{{\sinh \beta _1 }}$ and $\beta  = \beta _2<0 $ is the other sphere whose center is located on the negative $z$-axis at point $z = c~\coth \beta _2$ with radius of $r_{c2}=c/{{|\sinh \beta _2| }}$ (see Fig.~\ref{bis2} for details).  The parameter $c$ is non-dimensional and can be related to the separation distance $d$ (for two equal spheres) according to,
 \begin{eqnarray}
{\beta _1} =  - {\beta _2} = \cosh^{ - 1}\left( {\frac{d }{{2a}}} \right),
  \end{eqnarray}
\begin{eqnarray}
c = \sinh {\beta _1}.
 \end{eqnarray}
 The general three-dimensional solution of solute distribution in bispherical coordinate can be found via the following \emph{ansatz},
\begin{align}
n(\alpha ,\beta ,\phi ) &= \sqrt {\cosh \beta  - \cos \alpha } \sum\limits_{m = 0}^\infty  {\sum\limits_{n = m}^\infty  {\left[ {({{\tilde A}_{nm}}\cosh (n + 0.5)\beta  + {{\tilde B}_{nm}}\sinh (n + 0.5)\beta )\cos m\phi } \right.} }\nonumber \\
&\left. { + ({{\tilde C}_{nm}}\cosh (n + 0.5)\beta  + {{\tilde D}_{nm}}\sinh (n + 0.5)\beta )\sin m\phi } \right]P_n^m(\cos \alpha ),\label{laplacian}
\end{align}
where ${{\tilde A}_{nm}}$, ${{\tilde B}_{nm}}$, ${{\tilde C}_{nm}}$ and ${{\tilde D}_{nm}}$ are unknown coefficients that must be determined by applying the boundary conditions on particle surfaces and $P_n^m(\cos\alpha)$ is the associated Legendre function of the first kind. Having utilized the orthogonality of trigonometric functions and the associated Legendre polynomials, we have a set of recursive relationships which must be solved numerically. These relations are given in Appendix~A.\\
\subsection{Hydrodynamics}
 The hydrodynamics at Stokes regime in non-dimensional form can be written as

\begin{eqnarray}
 \nabla \cdot{\bf{v}} = 0,\label{hyd_1}
\end{eqnarray}
\begin{eqnarray}
 - \nabla p +{\nabla ^2}{\bf{v}} = 0.\label{hyd_2}
\end{eqnarray}
The above equations are then subjected to a slip boundary condition at the surfaces of the colloids ($j=1$ or $2$) and the quiescent far field condition while considering the fact that particles are force- and torque-free
\begin{eqnarray}
{\bf{v}}({\bf{r}} \in S_c^j) = {{\bf{v}}^j_s} + {\bf{U}}^j+{\bf{\Omega}}^j \times ({\bf{r}}-{\bf{r}}_0^j),
\end{eqnarray}
\begin{eqnarray}
{\bf{v}}(\left| {\bf{r}} \right| \to \infty ) \to 0,
\end{eqnarray}
\begin{eqnarray}
{{\bf{F}}}_T^j = \mathop{{\int\!\!\!\!\!\int}\mkern-21mu \bigcirc}\limits_{{S_c^j}} 
 {({\bf{e}}_n^j \cdot \sigma )dS = 0},\label{f_free}
\end{eqnarray}
\begin{eqnarray}
{{\bf{T}}}_T^j = \mathop{{\int\!\!\!\!\!\int}\mkern-21mu \bigcirc}\limits_{{S_c^j}} 
 {({\bf{r}} - {{\bf{r}}}_0^j) \times ({\bf{e}}_n^j \cdot \sigma )dS = 0},\label{t_free}
\end{eqnarray}
where $\sigma$ is the stress tensor, ${{\bf{r}}}_0^j$ is the location of the $j$th particle center of mass, ${\bf{U}}^j$ and ${\bf{\Omega}}^j$ are the non-dimensional translational and angular velocities of the $j$th colloid yet needed to be determined and finally ${{\bf{v}}^j_s}$ is the non-dimensional slip velocity on the surface of the $j$th colloidal motor, given as
\begin{eqnarray}
{{\bf{v}}_s^j} =  -{\nabla _s } n({\bf{r}} \in {S_c^j}),\label{slip}
\end{eqnarray}
where the surface gradient is defined as $\nabla _s= ({\bf{I}}-{\bf{e}}_n{\bf{e}}_n)\cdot \nabla$ and all variables introduced in the above are non-dimensionalized with $U_c \sim b{n}_c/{a}$, $\Omega_c\sim b{n}_c/{a^2}$,
$F_c \sim \mu U_c a$ and $T_c \sim \mu U_c a^2$. The mobility coefficient $b$ is define as
\begin{eqnarray}
b = \frac{{{k_B}T{L^2}}}{\mu }\int_0^\infty  {y'\left[ {\exp \left( { - \frac{\phi }{{{k_B}T}}} \right) - 1} \right]dy'},
\end{eqnarray}
where $\mu$ is the fluid viscosity, $k_B$ is the Boltzmann constant, $T$ is the absolute fluid temperature, $y'$ is the distance to the particle edge and $\phi$ is the net potential interaction of the solute molecules with the particle in the presence of solvent. For a hard-sphere excluded potential interaction between solutes and colloids, this reduces to $b = -k_BT L^2/{2\mu}$.

\begin{figure} 
\centering
\includegraphics[width=0.8 \textwidth]{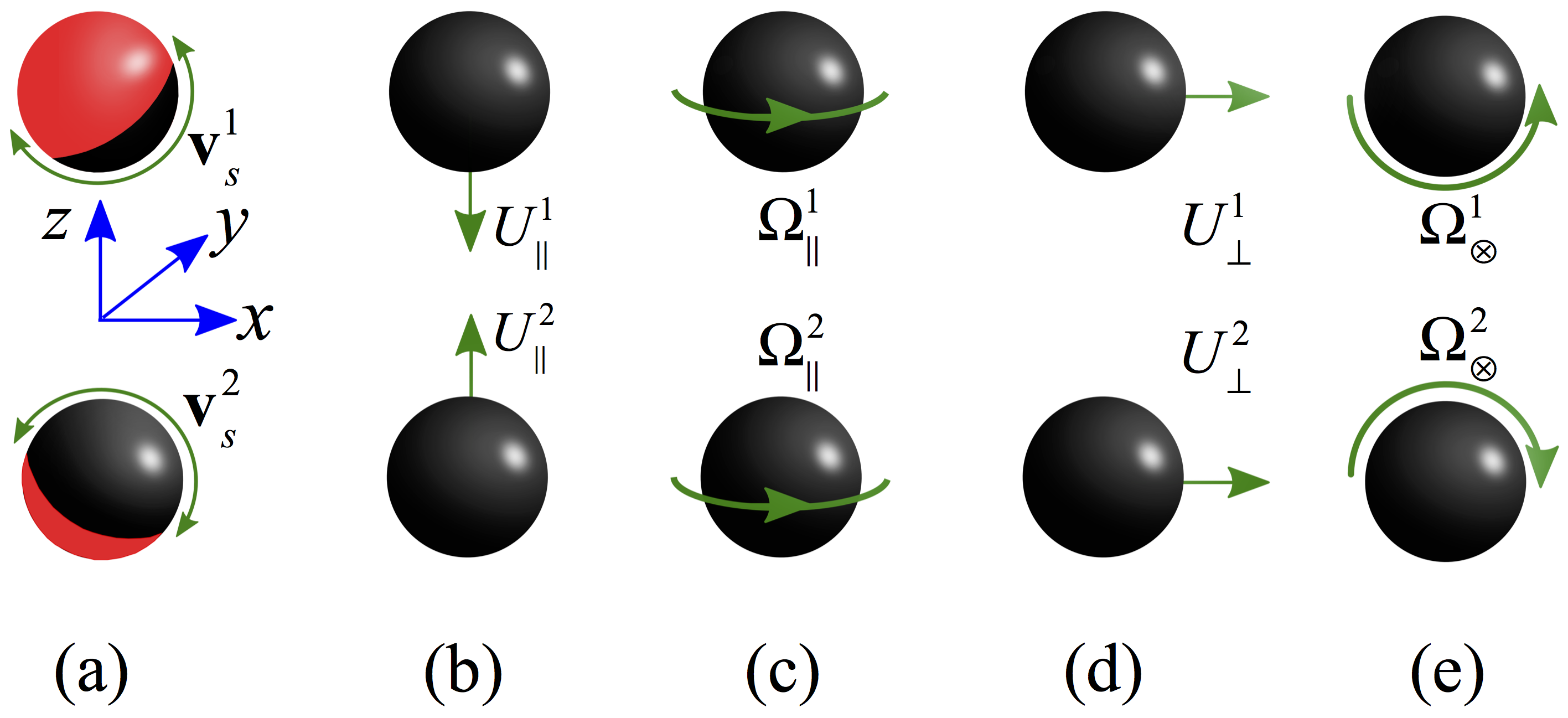}	     	    
 \caption{\footnotesize{Decomposition of velocity field into (a) surface diffusiophoretic slip with no net movement (translational or rotational) of the colloids (b) translation of colloids parallel to the line of centers ($z$ axis) with no rotation or slip (c) rotation of colloids around the line of centers ($z$ axis) with no slip or translation (d) translation of colloids perpendicular to the line of centers ($x$ or $y$ axis) with no slip or rotation} (e) rotation of colloids around the third axis ($y$ or $x$ axis) with no slip or translation.} 
\label{Fig_2}
 \end{figure}

The arbitrary motion of two colloidal particles in space can be decomposed into translational motion parallel $\parallel$ ($z$-axis) and perpendicular $\bot$ ($x$- and $y$-axes) to the line of centers and rotational motion around an axis parallel $\parallel$ ($z$-axis) to the line of centers and the third axis $\otimes$ ($y$- and $x$-axes), (see Fig.~\ref{Fig_2} for details). Furthermore, due to linearity of the Stokes flow equations and the boundary conditions, the fluid flow around two catalytically driven active particles can be decomposed into five subproblems (see Fig.~\ref{Fig_2}):\\
\begin{enumerate}
\item The colloidal particles are stationary, i.e. $U^{j}_{\parallel}=U^{j}_{\bot}=\Omega^{j}_{\parallel}=\Omega^{j}_{\otimes}=0$ and $j=1$ and $2$, but the fluid slips according to the diffusiophoretic slip (Eq.~\ref{slip}) on particles surfaces.\\
\item The colloidal particles are translating with velocities $U^{1}_{\parallel}$ and $U^{2}_{\parallel}$ parallel to the line of centers ($z$-direction) with no translations along the axis perpendicular to the line of centers ($U^{1}_{\bot}=U^{2}_{\bot}=0$) or rotations around any axes ($\Omega^{j}_{\parallel}=\Omega^{j}_{\otimes}=0$). In this case, the colloids experience hydrodynamic forces in the direction of motion $F^{j,H}_{\parallel}$, but the motion is axisymmetric and hence there are no hydrodynamics torques on particles.\\
\item The colloidal particles are rotating with angular velocities $\Omega^{1}_{\parallel}$ and $\Omega^{2}_{\parallel}$ around the axis parallel to the line of centers ($z$-direction) with no translations along any direction ($U^{j}_{\parallel}=U^{j}_{\bot}=0$) or rotations around any axes ($\Omega^{1}_{\otimes}=\Omega^{2}_{\otimes}=0$). In this case, the colloids experience hydrodynamic torques $T^{j,H}_{\parallel}$, but their motions are axisymmetric and hence there are no hydrodynamic forces on the particles. We will show below that this contribution is identically zero in self-diffusiophoresis.\\
\item The colloidal particles are translating with velocities $U^{1}_{\bot}$ and $U^{2}_{\bot}$ along an axis perpendicular to the line of centers with no translations along the axis parallel to the line of centers ($U^{1}_{\parallel}=U^{2}_{\parallel}=0$) or rotations around any axes ($\Omega^{j}_{\parallel}=\Omega^{j}_{\otimes}=0$). In this case, due to the hydrodynamic coupling between the motions of the particles, there are hydrodynamic forces in the direction of motion and torques around an axis along the third direction.\\ 
\item The colloidal particles are rotating with angular velocities $\Omega^{1}_{\otimes}$ and $\Omega^{2}_{\otimes}$ around the third axis with no translations along any axis ($U^{j}_{\parallel}=U^{j}_{\bot}=0$) or rotations around the axis parallel to the line of centers ($\Omega^{1}_{\parallel}=\Omega^{2}_{\parallel}=0$). In this case, due to the hydrodynamic coupling between the motions of the particles, there are hydrodynamic forces in a direction normal to the plane of translation.\\
\end{enumerate}
Note that in the above decomposition, each one of problems $(d)$ and $(e)$ are generally two problems (one along $x$ and the other along $y$-directions); nonetheless, the treatment for both are identical. To obtain the swimming velocities of the active colloids, the forces and torques associated with each one of the above five problems should be determined. Additionally, analytical solutions to the problems $(b)-(e)$ have already been addressed in the literature and we utilize these solutions in our analysis \cite{Jeffery,St1,Maude,Goldman,Wakiya,Davis,Majumdar,Spielman}.

To analyze the problem ($a$), the hydrodynamics (Eq.~\ref{hyd_1} and \ref{hyd_2}) must be solved. This can be achieved by utilizing the analytical solution for Stokes flow in bispherical coordinates \cite{JFM2}. In this case, the velocity components are in terms of infinite series with coefficients that must be found numerically. Thereafter, the forces and torques on the spheres associated with problem ($a$) can be numerically calculated. This method has been used in a number of studies, including ones involving the dynamics of two spherical particles, a spherical particle in the vicinity of a planar solid wall in electrophoresis \cite{Davis-Electro,Keh1} and the self-diffusiophoresis of an active particle near a solid wall \cite{Wall_1}. This approach is straight-forward, however, the details of the calculation and algebra are cumbersome and, hence, our objective here is to employ an alternative approach to obtain expressions for swimming translational and angular velocities ${\bf{U}}^j$ and ${\bf{\Omega}}^j$ of the particles due to self-diffusiophoresis. Therefore, we use the Reynolds Reciprocal Theorem (RRT) \cite{hapbren83,Miloh,Masood} to simplify the calculation of forces and torques on particles as explicit integrals in terms of the stress tensor of classical Stokes flow problems involving translational and rotational motion of two spherical particles. A significant advantage of RRT is that the hydrodynamics (Eqs.~\ref{hyd_1} and \ref{hyd_2}) need not be solved in order to evaluate the force and torque and therefore the number of calculations is significantly reduced..

Suppose (${\bf{v}}^{\prime},\sigma^{\prime}$) and (${{\bf{v''}}},\sigma''$) are the solutions of the Stokes flow of an incompressible fluid for some arbitrary fluid volume $\Sigma$. According to RRT, one can state that:
 \begin{eqnarray}
 \mathop{{\int\!\!\!\!\!\int}\mkern-21mu \bigcirc}\limits_{\partial \Sigma } 
 {{{\bf{n}}_{\Sigma}} \cdot \sigma ' \cdot {\bf{v''}}dS}  = \mathop{{\int\!\!\!\!\!\int}\mkern-21mu \bigcirc}\limits_{\partial \Sigma } 
 {{{\bf{n}}_{\Sigma}} \cdot \sigma '' \cdot {\bf{v'}}dS},\label{rrt}
 \end{eqnarray}
where $\partial \Sigma$ represents the boundaries enclosing the volume $\Sigma$; for instance, in the case of two spherical colloids, $\partial \Sigma$ accounts for a union of the boundaries including the colloids surfaces and an enclosing surface boundary far away from the two colloids, and ${\bf{n}}_{\Sigma}$ are the unit normals on these boundaries. Inasmuch as the dipolar field around a colloidal particle decays at least as ${\bf{v}}\sim r^{-2}$ and $\sigma \sim r^{-3}$, the far field area integral does not contribute to the calculation and can be neglected.

To evaluate the force and torque associated with problem ($a$) which hereinafter we call propulsive force and torque denoted by ${\bf{F}}^{j,P}$ and ${\bf{T}}^{j,P}$, respectively, we take advantage of the detailed solutions of four fundamental problems explained in the literature (see below for details) in conjunction with RRT. This can be achieved by first decomposing the force and torque on colloids in problem ($a$) in the parallel and perpendicular directions to the line of centers and then choosing appropriate test problems (${\bf{v''}},\sigma''$). Using this method, we can find the propulsive forces and torques on colloidal particles as integrals which can then be calculated numerically. Below, we explain these procedures in detail.
\subsubsection{Propulsive forces on two colloidal motors along the center-to-center line}
To quantify the forces on the particles along the line of centers ($z$-direction) in problem $(a)$ (Fig.~\ref{Fig_2}(a)), ($\bf{v}''$, $\sigma''$) in Eq.~\ref{rrt} was chosen to be the solution of translational motion of two colloids  along their line of centers which satisfy the following boundary conditions at particles surfaces and far field,
 \begin{eqnarray}
 {\bf{v''}}({\bf{r}} \in  S_c^j) = {{\bf{e}}_\parallel},~~(j = 1,2),
 \end{eqnarray}
   \begin{eqnarray}
 {\bf{v''}}(|{\bf{r}}| \to  \infty) = 0,
 \end{eqnarray}
 where $\bf{e}_{\parallel}$ is the unit normal vector along the line of centers. This problem was analyzed in detail by  Stimson and Jeffery \cite{St1}. Having substituted ($\bf{v}''$, $\sigma''$) in Eq.~\ref{rrt} it can be shown that
\begin{eqnarray}
F_\parallel^{1,P} + F_\parallel^{2,P} = \mathop{{\int\!\!\!\!\!\int}\mkern-21mu \bigcirc}\limits_{S_c^1} 
 {{{\bf{e}}_\beta } \cdot {{\bf{\sigma}} _{S - J}} \cdot {\nabla _s }n~dS}  - \mathop{{\int\!\!\!\!\!\int}\mkern-21mu \bigcirc}\limits_{S_c^2} 
 {{{\bf{e}}_\beta } \cdot {{\bf{\sigma}}_{S - J}} \cdot {\nabla _s }n~dS},\label{S-J}
\end{eqnarray}
where ${{\bf{\sigma}}_{S - J}}$ is the stress tensor corresponding to Stimson and Jeffery's analysis \cite{St1} and $F_z^{j,P}$ ($j=1$, $2$) can be given as
\begin{align}
&F_\parallel^{1,P} = - \mathop{{\int\!\!\!\!\!\int}\mkern-21mu \bigcirc}\limits_{S_c^1} 
 {{{\bf{e}}_\beta } \cdot \sigma ' \cdot {{\bf{e}}_\parallel}~dS}, \\
&F_\parallel^{2,P} = \mathop{{\int\!\!\!\!\!\int}\mkern-21mu \bigcirc}\limits_{S_c^2} 
 {{{\bf{e}}_\beta } \cdot \sigma' \cdot {{\bf{e}}_\parallel}~dS}. 
\end{align}
At the moment we have one equation (\ref{S-J}) with two unknowns ($F_\parallel^{1,P}$ and $F_\parallel^{2,P}$) and hence one more equation is required to close up the problem. To obtain the second relationship, RRT can be used one more time by assuming ($\bf{v}''$, $\sigma''$) in Eq.~\ref{rrt} is the solution of translational motion of two colloids along their line of centers which translate in an opposite direction, 
\begin{eqnarray}
 {\bf{v''}}({\bf{r}} \in  S_c^j) = \mp {{\bf{e}}_\parallel},~~(j = 1,2),\label{mp1}
 \end{eqnarray}
\begin{eqnarray}
 {\bf{v''}}(|{\bf{r}}| \to  \infty) = 0,
 \end{eqnarray}
where the minus sign in Eq.~\ref{mp1} is for particle $1$ on top and the plus sign is for the particle $2$ on the bottom. This problem was analyzed initially by Maude \cite{Maude} (but with some errors) and, by substitution of the solution, we have
 \begin{eqnarray}
   -F_\parallel^{1,P} + F_\parallel^{2,P} = \mathop{{\int\!\!\!\!\!\int}\mkern-21mu \bigcirc}\limits_{S_c^1} 
 {{{\bf{e}}_\beta } \cdot {\sigma _M} \cdot {\nabla _s }n~dS}  - \mathop{{\int\!\!\!\!\!\int}\mkern-21mu \bigcirc}\limits_{S_c^2} 
 {{{\bf{e}}_\beta } \cdot {\sigma _M} \cdot {\nabla _s }n~dS},\label{maude}
 \end{eqnarray}
  where ${{\bf{\sigma}}_{M}}$ is the stress tensor corresponding to Maude's analysis \cite{Maude}. Solving Eqs.~\ref{S-J} and \ref{maude} for the two unknowns $F_z^{1,P}$ and $F_z^{2,P}$ yields 
\begin{align}
F_\parallel^{1,P} = \mathop{{\int\!\!\!\!\!\int}\mkern-21mu \bigcirc}\limits_{S_c^1} 
 {{{\bf{e}}_\beta } \cdot \left(\frac{{{\sigma _{S - J}} - {\sigma _M}}}{2}\right) \cdot {\nabla _s }n~dS}  - \mathop{{\int\!\!\!\!\!\int}\mkern-21mu \bigcirc}\limits_{S_c^2} 
 {{{\bf{e}}_\beta } \cdot \left(\frac{{{\sigma _{S - J}} - {\sigma _M}}}{2}\right) \cdot {\nabla _s }n~dS},\label{fz1} \\
F_\parallel^{2,P} = \mathop{{\int\!\!\!\!\!\int}\mkern-21mu \bigcirc}\limits_{S_c^1} 
 {{{\bf{e}}_\beta } \cdot \left(\frac{{{\sigma _{S - J}} + {\sigma _M}}}{2}\right) \cdot {\nabla _s }n~dS}  - \mathop{{\int\!\!\!\!\!\int}\mkern-21mu \bigcirc}\limits_{S_c^2} 
 {{{\bf{e}}_\beta } \cdot \left(\frac{{{\sigma _{S - J}} + {\sigma _M}}}{2}\right) \cdot {\nabla _s }n~dS}.\label{fz2} 
\end{align}
The above integrals were calculated numerically to find the magnitude of the propulsive forces along the line of centers ($z$-direction) of two colloids.

\subsubsection{Propulsive torques on two colloidal motors parallel to their line of centers}
Propulsive torques on colloids around the line of centers ($z$-axis) can be directly evaluated with RRT by letting ($\bf{v}''$, $\sigma''$) in (Eq.~\ref{rrt}) be the solution of rotations of two colloids around their line of centers. This analysis was performed by Jeffery \cite{Jeffery} and it is recapitulated in Appendix~B. In this case, the problem is axisymmetric and the only non-zero component of the velocity field is in the azimuthal angle direction $\phi$, and hence all non-zero components of the stress tensor associated with this problem do not have $\phi$ dependency and the propulsive torques on colloids along the line of centers can be proven to be identically zero (orthogonality of $\sin \phi$ and $\cos\phi$ functions),
\begin{eqnarray}
{T}_\parallel ^{j,P} = \mp \mathop{{\int\!\!\!\!\!\int}\mkern-21mu \bigcirc}\limits_{S_c^j} 
 {{{\bf{e}}_\beta } \cdot {\sigma _J} \cdot {\nabla _s}n\;dS}  = 0,
\end{eqnarray} 
where $\sigma _J$ is the non-dimensional stress tensor associated with Jeffery's problem \cite{Jeffery}.

\subsubsection{Propulsive forces and torques on two colloidal motors perpendicular to their line of centers and around the third axis}
In this case, the propulsive forces and torques on colloidal particles are in a direction perpendicular to the line of centers and around the third axis respectively. These directions can be either in $x$ or $y$ (see Fig.~\ref{Fig_1}(d) and (e)) and the analysis for each one of the directions is similar. The translational (or rotational) motion of colloidal particles in this case is no longer axisymmetric and the colloids experience a torque (or force) in the $y$ (or $x$) direction.

To use RRT (Eq.~\ref{rrt}) for calculation of the propulsive forces and torques on the colloids, ($\bf{v}''$, $\sigma''$) is chosen to be the solution of four sub-problems:\\

\noindent $(I)$ Two colloids are translating with no rotations in the same direction with a unit velocity perpendicular to the  line of centers   
\begin{align}
&{\bf{v}''}({\bf{r}} \in S_c^1) = {{\bf{e}}_\bot},\\
&{\bf{v}''}({\bf{r}} \in S_c^2) = {{\bf{e}}_\bot},\\
&{\bf{v}''}(|{\bf{r}}| \to \infty ) = 0,
\end{align}
where ${{\bf{e}}_\bot}$ can be either ${{\bf{e}}_x}$ or ${{\bf{e}}_y}$. This problem was analyzed by Goldman \emph{et~al.} \cite{Goldman}. Having utilized RRT (Eq.~\ref{rrt}), one can show
\begin{eqnarray}
 F_{\bot }^{1,P}+   F_\bot ^{2,P}=\mathop{{\int\!\!\!\!\!\int}\mkern-21mu \bigcirc}\limits_{S_c^1} 
 {{{\bf{e}}_{\beta} } \cdot {\tilde{\sigma _1}} \cdot {\nabla _s }n~dS}  - \mathop{{\int\!\!\!\!\!\int}\mkern-21mu \bigcirc}\limits_{S_c^2} 
 {{{\bf{e}}_\beta } \cdot {\tilde{\sigma _1}} \cdot {\nabla _s }n~dS},\label{br_1}
 \end{eqnarray}
where $\tilde{\sigma}_{1}$ is the non-dimensional stress tensor corresponding to analysis in \cite{Goldman}.\\
$(II)$~Two colloids are rotating with no translation in opposite directions around the third direction which can be either the $y$- or $x$-axis,   
\begin{align}
&{\bf{v}''}({\bf{r}} \in S_c^1) = {{\bf{e}}_{\otimes}} \times ({\bf{r}} - {{\bf{r}}}_0^1),\\
&{\bf{v}''}({\bf{r}} \in S_c^2) = -{{\bf{e}}_{\otimes}} \times ({\bf{r}} - {{\bf{r}}}_0^2),\\
&{\bf{v}''}(|{\bf{r}}| \to \infty ) = 0,
\end{align}
where ${{\bf{e}}_{\otimes}}$ accounts for the unit vector in the third direction, i.e. $y$ or $x$. This problem was also analyzed by Goldman \emph{et~al.} \cite{Goldman}. Having utilized RRT (Eq.~\ref{rrt}), one can show
\begin{eqnarray}
 T_\otimes^{1,P}- T_\otimes^{2,P}  = \mathop{{\int\!\!\!\!\!\int}\mkern-21mu \bigcirc}\limits_{S_c^1} 
 {{{\bf{e}}_\beta } \cdot {\tilde{\sigma _2}} \cdot {\nabla _s}n~dS}  - \mathop{{\int\!\!\!\!\!\int}\mkern-21mu \bigcirc}\limits_{S_c^2} 
 {{{\bf{e}}_\beta } \cdot {\tilde{\sigma _2}} \cdot {\nabla _s}n~dS},\label{br_2}
 \end{eqnarray}
 where $\tilde{\sigma}_{2}$ is the non-dimensional stress tensor corresponding to the analysis performed by Goldman \emph{et~al.} analysis \cite{Goldman}.\\
$(III)$~Two colloids are translating with no rotation in an opposite direction with a unit velocity perpendicular to the line of centers,   
\begin{align}
&{\bf{v}''}\left( {{\bf{r}} \in S_c^1} \right)  = {{\bf{e}}_\bot },\\
&{\bf{v}''}\left( {{\bf{r}} \in S_c^2} \right)  = -{{\bf{e}}_\bot },\\
&{\bf{v}''}\left( {|{\bf{r}}| \to \infty } \right) = 0.
\end{align}
This problem was analyzed by O'Neill \cite{ONeill}. Having utilized RRT (Eq.~\ref{rrt}), we find
\begin{eqnarray}
 F_\bot ^{1,P}- F_\bot ^{2,P}  = \mathop{{\int\!\!\!\!\!\int}\mkern-21mu \bigcirc}\limits_{S_c^1} 
 {{{\bf{e}}_\beta } \cdot {\tilde{\sigma _3}} \cdot {\nabla _s}n~dS}  - \mathop{{\int\!\!\!\!\!\int}\mkern-21mu \bigcirc}\limits_{S_c^2} 
 {{{\bf{e}}_\beta } \cdot {\tilde{\sigma _3}} \cdot {\nabla _s}n~dS},\label{on_1}
 \end{eqnarray}
where $\tilde{\sigma}_{2}$ is the non-dimensional stress tensor corresponding to the analysis performed by O'Neill \cite{ONeill}.\\
$(IV)$~Two colloids are rotating with no translation in a same direction   
\begin{align}
&{\bf{v''}}\left( {{\bf{r}} \in S_c^1} \right) = {{\bf{e}}_ \otimes } \times \left( {{\bf{r}} - {\bf{r}}_0^1} \right),\\
&{\bf{v''}}\left( {{\bf{r}} \in S_c^2} \right) = {{\bf{e}}_ \otimes } \times \left( {{\bf{r}} - {\bf{r}}_0^2} \right),\\
&{\bf{v''}}\left( {|{\bf{r}}| \to \infty } \right) = 0.
\end{align}
This problem was also analyzed by O'Neill \cite{ONeill}. Having utilized RRT (Eq.~\ref{rrt}), we find
\begin{eqnarray}
 T_\otimes^{1,P}+T_\otimes^{2,P}  = \mathop{{\int\!\!\!\!\!\int}\mkern-21mu \bigcirc}\limits_{S_c^1} 
 {{{\bf{e}}_\beta } \cdot {\tilde{\sigma _4}} \cdot {\nabla _s}n~dS}  - \mathop{{\int\!\!\!\!\!\int}\mkern-21mu \bigcirc}\limits_{S_c^2} 
 {{{\bf{e}}_\beta } \cdot {\tilde{\sigma _4}} \cdot {\nabla _s}n~dS},\label{on_2}
 \end{eqnarray}
 where $\tilde{\sigma}_{2}$ is the non-dimensional stress tensor correspond to O'Neill's analysis\cite{ONeill}.
 
 By combining Eqs.~\ref{br_1} and \ref{on_1}, the forces on colloids are shown to be
 \begin{align}
&F_\bot ^{1,P} =   \mathop{{\int\!\!\!\!\!\int}\mkern-21mu \bigcirc}\limits_{S_c^1} 
 {{{\bf{e}}_\beta} \cdot \left(\frac{{{{\tilde \sigma }_1} + {{\tilde \sigma }_3}}}{2}\right) \cdot {\nabla _s}n~dS}  - \mathop{{\int\!\!\!\!\!\int}\mkern-21mu \bigcirc}\limits_{S_c^2} 
 {{{\bf{e}}_\beta} \cdot \left(\frac{{{{\tilde \sigma }_1} + {{\tilde \sigma }_3}}}{2}\right) \cdot {\nabla _s}n~dS}, \label{fx1}\\
&F_\bot^{2,P} =   \mathop{{\int\!\!\!\!\!\int}\mkern-21mu \bigcirc}\limits_{S_c^1} 
 {{{\bf{e}}_\beta} \cdot \left(\frac{{{{\tilde \sigma }_1} - {{\tilde \sigma }_3}}}{2}\right) \cdot {\nabla _s}n~dS} -  \mathop{{\int\!\!\!\!\!\int}\mkern-21mu \bigcirc}\limits_{S_c^2} 
 {{{\bf{e}}_\beta} \cdot \left(\frac{{{{\tilde \sigma }_1} - {{\tilde \sigma }_3}}}{2}\right) \cdot {\nabla _s}n~dS},\label{fx2} 
\end{align}
 and from Eqs.~\ref{br_2} and \ref{on_2}, the torques around the third axis can be written as
\begin{align}
&T_\otimes^{1,P} =   \mathop{{\int\!\!\!\!\!\int}\mkern-21mu \bigcirc}\limits_{S_c^1} 
 {{{\bf{e}}_\beta} \cdot \left(\frac{{{{\tilde \sigma }_4} + {{\tilde \sigma }_2}}}{2}\right) \cdot {\nabla _s}n~dS}  - \mathop{{\int\!\!\!\!\!\int}\mkern-21mu \bigcirc}\limits_{S_c^2} 
 {{{\bf{e}}_\beta} \cdot \left(\frac{{{{\tilde \sigma }_4} + {{\tilde \sigma }_2}}}{2}\right) \cdot {\nabla _s}n~dS}, \label{tx1}\\
&T_\otimes^{2,P} =   \mathop{{\int\!\!\!\!\!\int}\mkern-21mu \bigcirc}\limits_{S_c^1} 
 {{{\bf{e}}_\beta} \cdot \left(\frac{{{{\tilde \sigma }_4} - {{\tilde \sigma }_2}}}{2}\right) \cdot {\nabla _s}n~dS}  - \mathop{{\int\!\!\!\!\!\int}\mkern-21mu \bigcirc}\limits_{S_c^2} 
 {{{\bf{e}}_\beta} \cdot \left(\frac{{{{\tilde \sigma }_4} - {{\tilde \sigma }_2}}}{2}\right) \cdot {\nabla _s}n~dS},\label{tx2} 
\end{align} 
where the force and torques on the particles are
\begin{align}
F_ \bot ^{1,P} &= -\mathop{{\int\!\!\!\!\!\int}\mkern-21mu \bigcirc}\limits_{S_c^1} 
 {{{\bf{e}}_\beta } \cdot \sigma ' \cdot {{\bf{e}}_ \bot }~dS} ,\\
F_ \bot ^{2,P} &=   \mathop{{\int\!\!\!\!\!\int}\mkern-21mu \bigcirc}\limits_{S_c^2} 
 {{{\bf{e}}_\beta } \cdot \sigma ' \cdot {{\bf{e}}_ \bot }~dS} ,\\
T_ \otimes ^{1,P} &=  - \mathop{{\int\!\!\!\!\!\int}\mkern-21mu \bigcirc}\limits_{S_c^1} 
 {\left( {{\bf{r}} - {\bf{r}}_0^1} \right) \times \left( {{{\bf{e}}_\beta } \cdot \sigma '} \right) \cdot {{\bf{e}}_ \otimes }~dS},\\
T_ \otimes ^{2,P} &=  \mathop{{\int\!\!\!\!\!\int}\mkern-21mu \bigcirc}\limits_{S_c^2} 
 {\left( {{\bf{r}} - {\bf{r}}_0^2} \right) \times \left( {{{\bf{e}}_\beta } \cdot \sigma '} \right) \cdot {{\bf{e}}_ \otimes }~dS}.
\end{align}
The solution to the four classical problems (cite{Goldman,ONeill} mentioned above are recapitulated in Appendix~B. The above relations, Eqs.~\ref{fx1} -- \ref{tx2}, along with Eqs.~\ref{fz1} and \ref{fz2} determine the propulsive forces and torques on the particles in problem $(a)$. The integrands of these integrals were all evaluated according to the analytical relations, however the integrations were performed numerically.  In the evaluation of these quadratures, all stress tensors associated with the classical problems are in the form of infinite sums and, in our calculation, these series are truncated according to the fact that the coefficients vanish for a sufficiently large number of terms. Also, more terms in the truncated series were required for small separation distances to reach numerically accurate and consistent results for both concentration field and stress tensor components.

\subsection{Swimming translational and angular velocities}

To capture the swimming translational and angular velocities, we take advantage of the facts that the colloidal particles are force and torque free (Eqs.~\ref{f_free} -- \ref{t_free}) and hence the propulsive force and torque in problem ($a$) should be balanced by the hydrodynamic forces and torques associated with problems $(b)$ -- $(e)$ on each colloid (see Fig.~\ref{Fig_2}).
For the motions of colloids in arbitrary directions, we have
 \begin{align}
&F_{\parallel}^{j,T} =  F_{\parallel }^{j,H} + F_{\parallel }^{j,P}= 0,\label{FTz}\\
&F_{\bot}^{j,T} =  F_{\bot }^{j,H} + F_{\bot }^{j,P}= 0,\label{FxT}\\
&T_{\otimes}^{j,T} =  T_{\otimes }^{j,H} + T_{\otimes }^{j,P}= 0,\label{TyT}
\end{align}
where each one of the above equations must be written for the $j$th colloid ($j=1~\rm{or}~2$) and Eqs.~\ref{FxT} and \ref{TyT} are in two directions ($x$ and $y$-directions in our convention, see Fig.~\ref{Fig_2}). Notice that since there are no propulsive torques on particles around the axis parallel to the line of centers, the particles do not rotate around this axis $\Omega^{1}_{\parallel}=\Omega^{2}_{\parallel}=0$.

The hydrodynamic force on the $j$th particle $F_{\parallel }^{j,H}$ is found by direct analysis of the slow viscous motions of two colloids with uniform velocities of $U_1$ and $U_2$ in an otherwise quiescent fluid. The general solution of Stokes motion of two particles with arbitrary radii and identical velocities along their line of centers  was first proposed by Stimson and Jeffery \cite{St1}. Later, Maude \cite{Maude} utilized a similar approach (but with some errors) for two particles  translating with equal and opposite velocities along their line of centers. The general solution for the motion of two particles with arbitrary radii and velocities is revealed by superposition of these two analyses. Spielman \cite{Spielman} provided the analysis and corrected relations for the motion of two particles with arbitrary radii and velocities and it was shown that the hydrodynamic forces on two particles along the line of centers ($z$-direction) can be written as
\begin{align}
F_{\parallel}^{1,H} =  - {\kappa _1}{U_{\parallel}^1} + {\lambda _1}{U_{\parallel}^2},\label{fpzz}\\
F_{\parallel}^{2,H} =  - {\kappa _2}{U_{\parallel}^2} + {\lambda _2}{U_{\parallel}^1},
\end{align}
where $\kappa_1$, $\kappa_2$, $\lambda_1$ and $\lambda_2$ are positive coefficients which solely depends on particles radii and their separation distance $\Delta$.

The hydrodynamic forces and torques exerted on the particles due to their translational and rotational motions along the perpendicular to the line of centers $\bf{e}_\bot$ and the third axes $\bf{e}_\otimes$, respectively, are found via superposition of four subproblems which are discussed in Appendix~B. Generally, any translational motion of two particles with arbitrary velocities $U_ \bot ^1$ and $U_ \bot ^2$ can be decomposed into two problems: (i) The particles are translating with similar velocities of $(U_ \bot ^1+U_ \bot ^2)/{2}$ in the same direction perpendicular to the line of centers and (ii) the particles are translating with similar velocities $({U_ \bot ^1-U_ \bot ^2})/{2}$ in opposite directions perpendicular to the line of centers. Likewise, similar decompositions can be formulated for rotational motions of the colloids around the third axis. This leads us to have the hydrodynamic forces and torques exerted on the particles as linear superpositions of these four problems and represent it in the following form
\begin{eqnarray}
\mathcal{F} = \mathcal{R} \cdot \mathcal{U},\label{fnxx}
\end{eqnarray}
where
\begin{eqnarray}
\mathcal{F} = \left[ {\begin{array}{*{20}{c}}
{F_ \bot ^1}\\
{F_ \bot ^2}\\
{T_ \otimes ^1}\\
{T_ \otimes ^2}
\end{array}} \right],~~~~~
\mathcal{U} = \left[ {\begin{array}{*{20}{c}}
{U_ \bot ^1}\\
{U_ \bot ^2}\\
{\Omega _ \otimes ^1}\\
{\Omega _ \otimes ^2}
\end{array}} \right],
\end{eqnarray}
and $\mathcal{R}$ is the resistance matrix with components found from linear combinations of four problems discussed  above. By substituting the hydrodynamic forces and torques on the colloids given in Eqs.~\ref{fpzz} -- \ref{fnxx} and the propulsive forces and torques given in Eqs.~\ref{fz1}, \ref{fz2} and \ref{fx1} -- \ref{tx2} into the force and torque balances in Eqs.~\ref{FTz} -- \ref{TyT}, we can determine the swimming translational velocities ${\bf{U}}^{j}$ and angular velocities ${\bf{\Omega}}^{j}$ of the two colloidal motors. 

\section{Results and discussions}\label{S2}
At a given separation distance, the dynamics of two catalytic motors depends on colloids coverage $\theta^{j}_{cap}$ and their relative orientation ${\bf{n}}^{j}_{P}$. Here, we consider the case where the orientation vectors are in the same plane and we choose this to be the $x$ -- $z$ plane. Furthermore, a net repulsive hard-sphere excluded volume interaction is considered between product solute and colloids in the presence of solvent so that the particles swim along their orientation vectors when they are far away from each other. 
\subsection{Concentration field}
\begin{figure} 
\centering
\includegraphics[width=0.8 \textwidth]{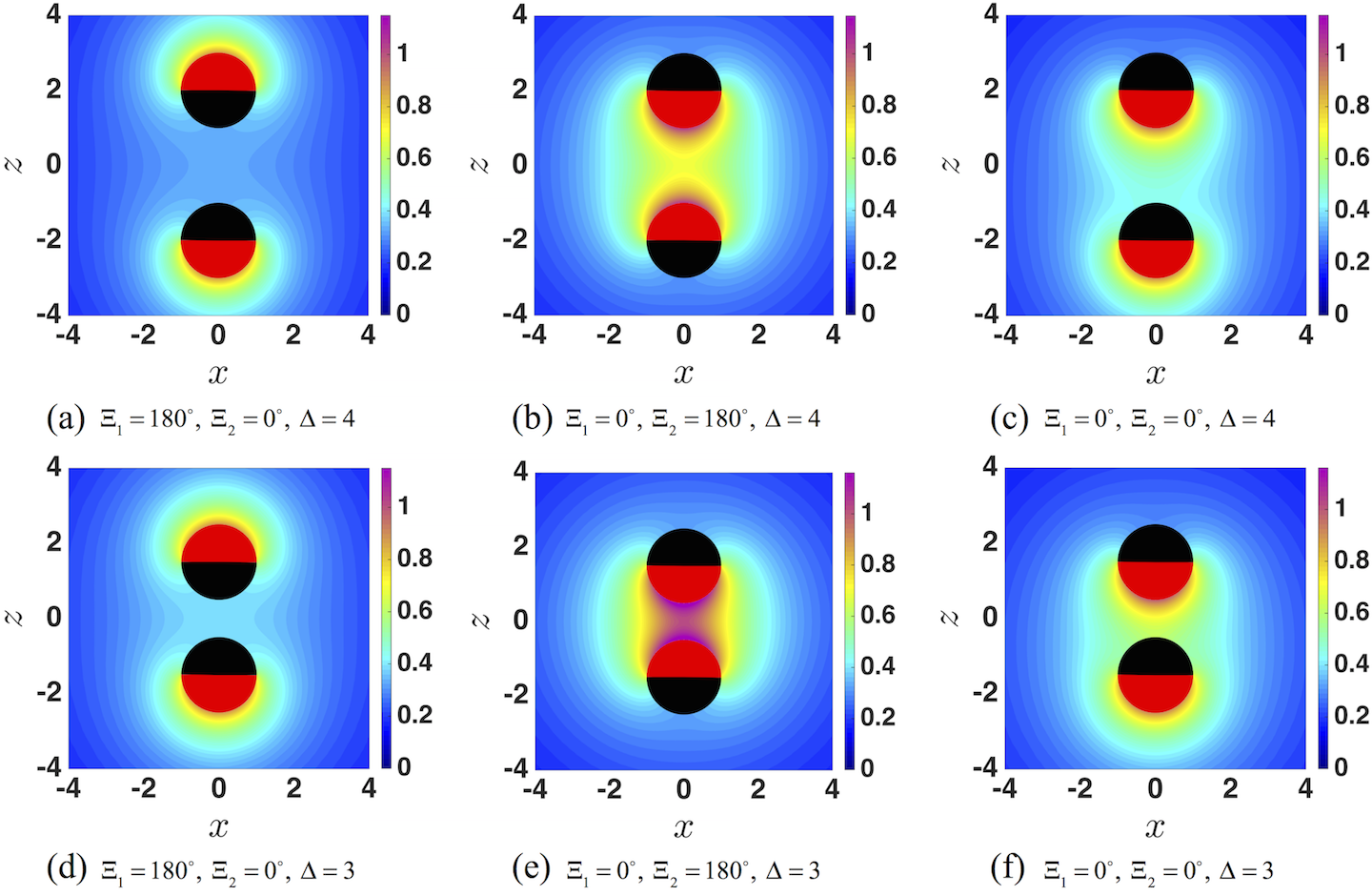}	     	    
 \caption{Non-dimensional concentration field for half active Janus colloids $(\theta^1 _{cap}=\theta^2 _{cap}  = 90^ \circ)$ in the $x$--$z$ plane for axisymmetric orientation of colloids with $\Xi_1  = 180^ \circ$ and $\Xi_2  = 0^ \circ$ at two different non-dimensional center-to-center distances of $(a)$ $\Delta=4$ and $(b)$ $\Delta=3$ and $\Xi_1  = 0^ \circ$ and $\Xi_2  = 180^ \circ$ at two different non-dimensional center-to-center distances of $(c)$ $\Delta=4$ and $(d)$ $\Delta=3$. Notice the red caps indicate the catalytic surface of the colloids.} 
\label{Fig_3}
 \end{figure}

The concentration field around the colloids is independent of azimuthal angle $\phi$ when the inclination angles of the colloids are either $0^ \circ$ or $180^ \circ$. For the situation where the active sections of colloids face opposite to each other, i.e. $\Xi_1  = 180^ \circ$ and $\Xi_2  = 0^ \circ$, the solute distribution does not change drastically, except for very high coverage, as the colloids approach each other since the solutes can freely diffuse away from them (see Fig.~\ref{Fig_3}(a) and (d)). On the contrary, in the other case where the active caps of the colloids face toward each other, $\Xi_1  = 0^ \circ$ and $\Xi_2  = 180^ \circ$, as the separation distance between the particle decreases, the concentration of solute increases precipitously in the gap region adjacent to the colloids (see Fig.~\ref{Fig_3}(b) and (e)). 
When the active section of one colloid faces the passive section of the other one, i.e. $\Xi_1 = \Xi_2  = 0^ \circ$, the concentration of solute in the gap increases again as the separation distance between the colloids decreases. The major difference between this situation and the other two previous axisymmetric cases is the fact that the concentration field is no longer symmetric with respect to the mid-plane $x$ -- $y$ and hence, as we show later, the swimming velocities of the colloids are no longer equal in magnitudes (see Fig.~\ref{Fig_3}(c) and (f)).

\begin{figure} 
\centering
\includegraphics[width=0.8 \textwidth]{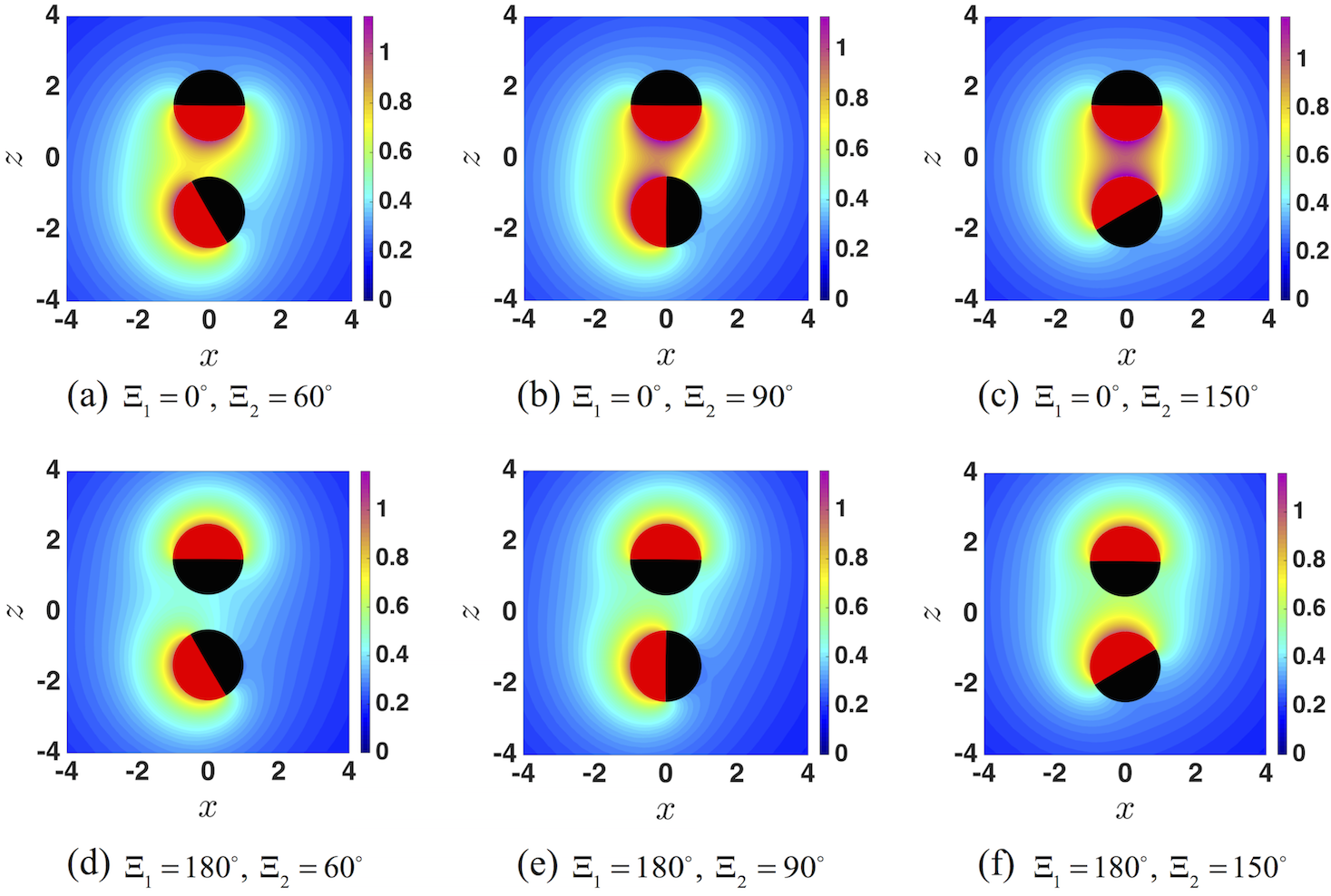}	     	    
 \caption{Non-dimensional concentration field for half active Janus colloids $(\theta^1 _{cap}=\theta^2 _{cap}  = 90^ \circ)$ in the $x$-$z$ plane for asymmetric orientation and non-dimensional center-to-center distances of $\Delta=3$. Notice the red caps indicate the catalytic surface of the colloids.} 
\label{Fig_4}
 \end{figure}

For all other orientations, the solution of the concentration field is three dimensional in bispherical coordinates, $n\left( {\alpha ,\beta ,\phi } \right)$. Representative concentration fields in the $x$ -- $z$ plane around two Janus colloids $(\theta^1 _{cap} =\theta^2 _{cap} = 90)$ for various relative inclination angles and a fixed separation distance $\Delta=3$  are shown in Fig. \ref{Fig_4}. The pair interaction effect is more pronounced when the active sections are close to each other and therefore the solute concentration is boosted in the gap. Furthermore, for all values of the inclination angles except the symmetric cases, the concentration fields around the colloids are no longer symmetric with respect to center-to-center line (see Fig.~\ref{Fig_2}) and hence results in the rotation of particles. These rotations are the direct consequence of diffusiophoresis which is due to the asymmetric distribution of product solute; this effect will be discussed in detail presently. Furthermore, the asymmetric distribution around the colloid axis is more pronounced as the inclination angles of colloids approach $\Xi_j  \to 90^ \circ$ or $\Xi_j  \to 270^ \circ$ ($j=1,2$). 

\begin{figure} 
\centering
\includegraphics[width=0.8 \textwidth]{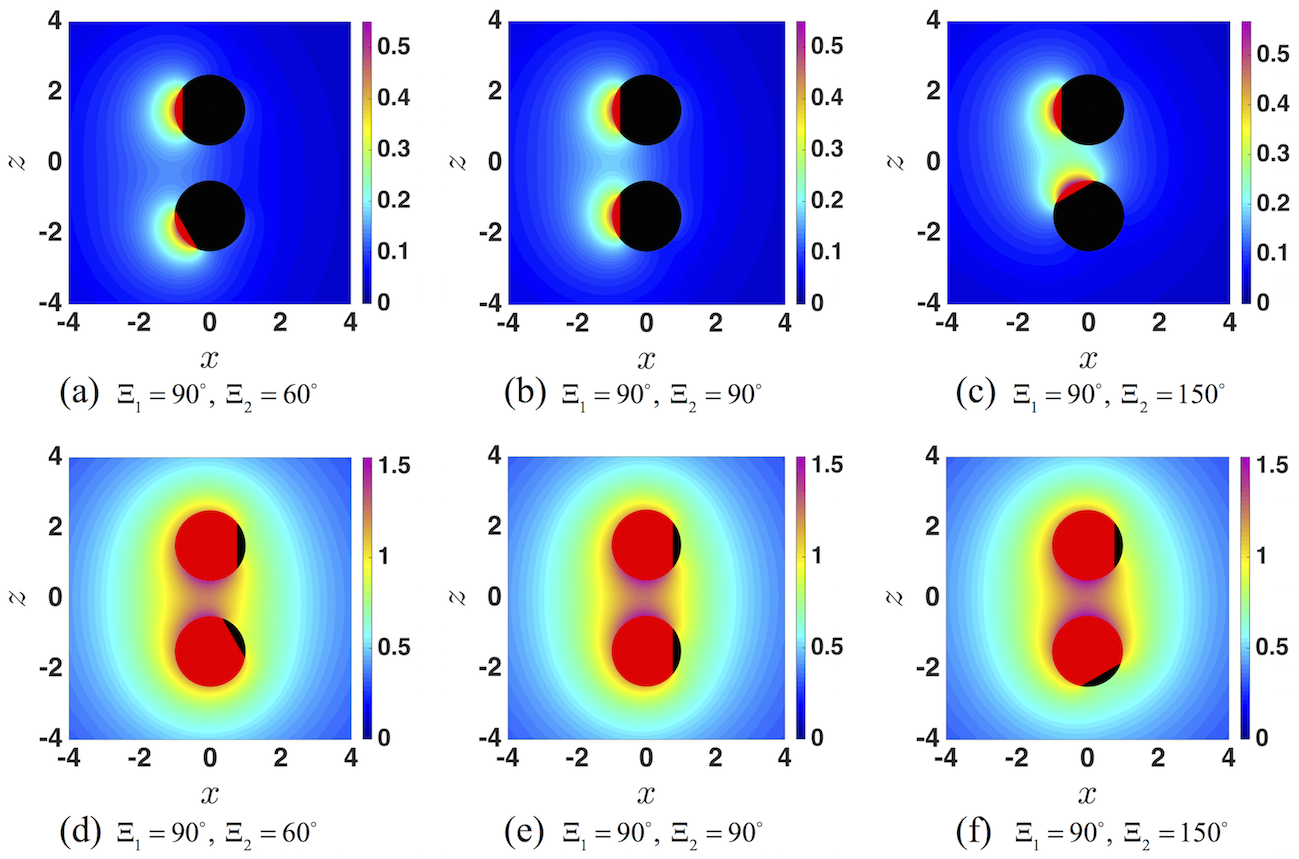}	     	    
 \caption{Non-dimensional concentration field for active colloids at various inclination angles in the $x$-$z$ plane for asymmetric orientation of colloids with $\Xi_1=\Xi_2  = 0^ \circ$ at  non-dimensional center-to-center distances of  $\Delta=3$ with $(a)-(c)$ $\theta^1 _{cap}=\theta^2 _{cap}  = 40^ \circ$ and $(d)-(f)$ $\theta^1 _{cap}=\theta^2 _{cap}  = 140^ \circ$. Notice the red caps indicate the catalytic surface of the colloids.} 
\label{Fig_5}
 \end{figure}

The concentration field around colloids can be altered not only by changing the orientation of colloids but also by varying the extent of active sections. Fig.~\ref{Fig_5}(a) -- (c) demonstrate concentration fields in the $x-z$ plane for small  $\theta _{cap}^1  =\theta _{cap}^2= 40^ \circ$ and large ($\theta _{cap}^1  =\theta _{cap}^2= 140^ \circ)$ active coverages. Note that for $\theta _{cap}^j  = 140^ \circ$, there is a substantial increase in solute concentration in the gap relative to the case where $\theta _{cap}^j  = 40^ \circ$ for similar inclination angles. Thus, the diffusiophoretic effect for colloids with large active coverages is more significant compared to the colloids with smaller coverages.
 
 \subsection{Swimming velocities and trajectories}
\subsubsection{ Axisymmetric motion of two colloidal motors swimming in the direction parallel to the center-to-center line}
To study the hydrodynamics of colloidal self-propulsion, we first focus on the axisymmetric motion of two motors, i.e. the inclination angles of the colloids are either $0^ \circ$ or $180^ \circ$. In this situation, the only non-zero component of the  swimming velocity is in the $z$-direction (see Fig.~\ref{Fig_1}). Moreover, since the flow is axisymmetric, the colloids do not undergo a rotation around any axis. This condition simplifies the governing equations significantly and the flow field can be found easily by using the conventional stream function technique in bispherical coordinates for the Stokes flow regime  \cite{Popescu1,Nima_thesis,Kapral2}. Additionally, to obtain a far field solution, we employ a method of reflections to validate our analytical solution. The detailed derivation of this solution is discussed in Appendix~C.
\begin{figure} 
\centering
\includegraphics[width=0.9 \textwidth]{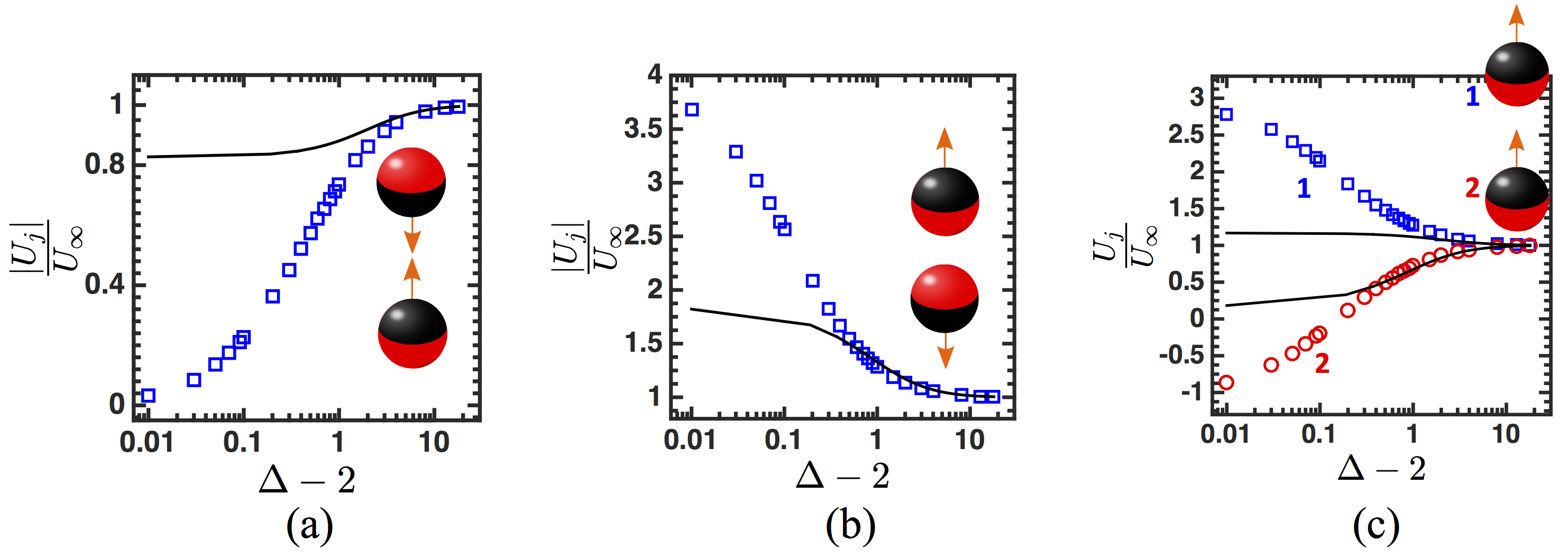}	     	    
 \caption{Non-dimensional swimming velocities of two Janus colloidal motors ($\theta^1 _{cap}=\theta^2 _{cap}  = 90^ \circ$) as a function of non-dimesnional separation distance $\Delta-2$ for axisymmetric motions. Open symbols represents the full solution while the solid lines are the results from the method of reflections.} 
\label{Fig_6}
 \end{figure}

\begin{figure} 
\centering
\includegraphics[width=0.7 \textwidth]{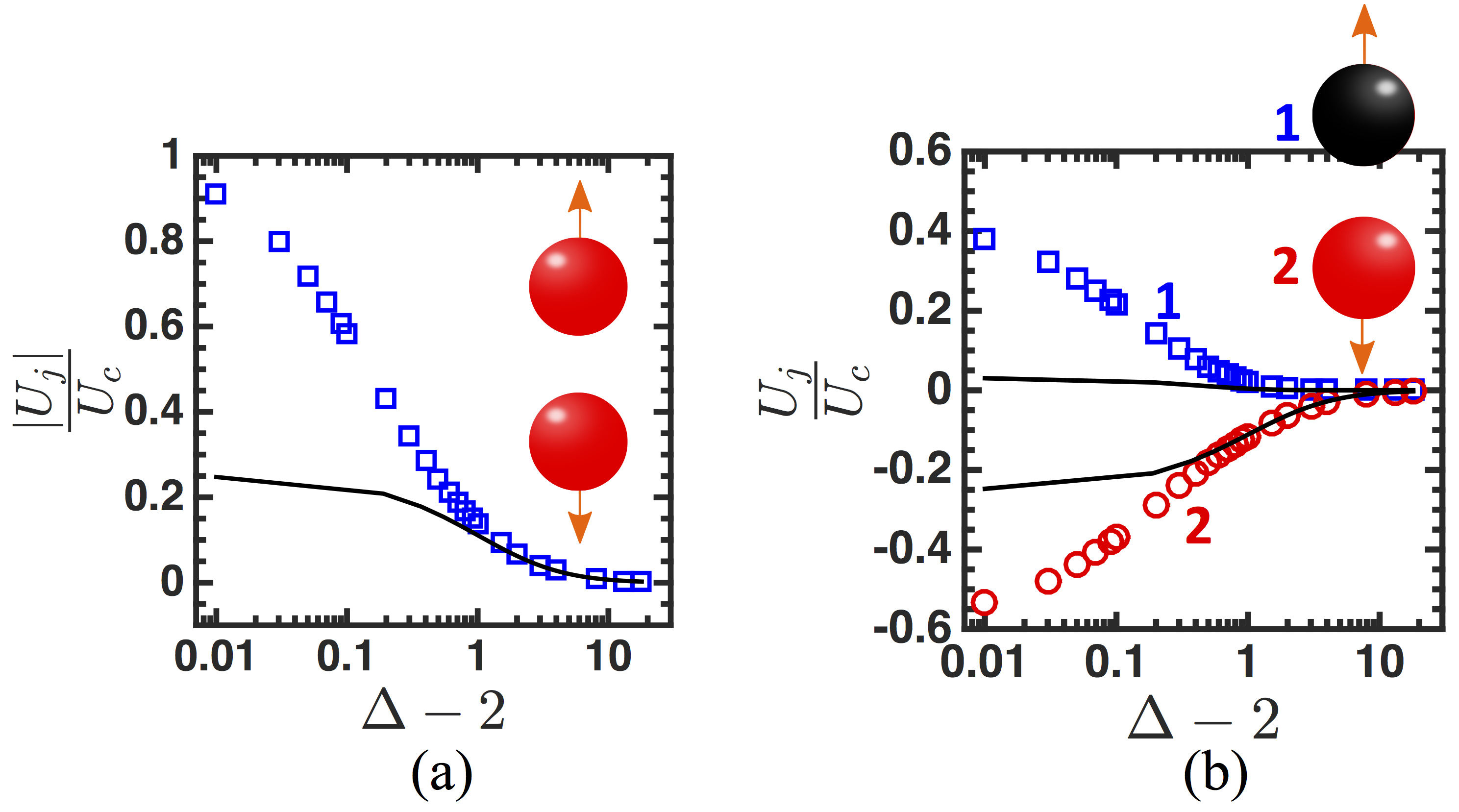}	     	    
 \caption{Non-dimensional swimming velocities of (a) two catalytically active colloids and (b) a catalytically active colloid and an inert colloid (cargo). Solid black lines are the method of reflections solutions.} 
\label{Fig_7}
 \end{figure}

Fig.~\ref{Fig_6} illustrates the non-dimensional swimming velocities of two Janus colloids as a function of the center-to-center separation distance $\Delta$ for three relative orientations. In all cases, the colloids move along the center-to-center line ($z$-direction), however, the dependency of the swimming velocities on the separation distance are substantially different for them. The concentration field can be symmetric or asymmetric with respect to the $x$ -- $y$ plane. For a situation where the active sections of the motors face opposite to each other, Fig.~\ref{Fig_6}(a), the swimming velocities of the colloids are equal in magnitude due to the symmetric orientation of the colloids with respect to the $x$ -- $y$ plane, and the magnitudes of swimming velocities of the colloids are equal to the reference swimming velocity $U_\infty$ when the colloids are a few radii away from each other. This magnitude decreases monotonically as a function of separation distance between the colloids. In this case, the propulsive force does not change as the particles approach each other because the concentration field around each one of them does not change drastically by the presence of the other (see Fig.~\ref{Fig_4}(a) and (d)) and therefore the reduction in swimming velocity is solely due to an increase in hydrodynamic resistance which scales as $\sim 1/({\Delta-2})$ \cite{Kim}. In contrast, when the active sections of the colloids face toward each other, the swimming velocities of particles increase considerably as the separation distance between them becomes smaller as shown in Fig.~\ref{Fig_6}(b). In this case, the concentration of solute in the gap between the colloids becomes larger in magnitude precipitously as the separation distance between the colloids shrinks (see Figs.~\ref{Fig_4}(b) and (e)) and consequently the propulsive force is greatly amplified so that it overtakes the increase in lubrication resistance. When the relative orientation of the colloids is not symmetric with respect to the $x$ -- $y$ plane, e.g. the active section of one colloid faces the passive section of the other colloid, the swimming velocities of the colloidal particles are no longer identical in magnitude because the concentration field around the colloids are no longer symmetric with respect to $x$ -- $y$ plane. Here, for a given orientation in Fig.~\ref{Fig_4}(c), as the separation distance decreases, the concentration gradient around the colloid on top (colloid $1$) increases while the concentration gradient around the one on the bottom (colloid $2$) decreases and hence they swim faster and slower, respectively. At very small separation distances, $\Delta-2\sim 0.3$, the concentration build up in the thin gap becomes so strong that it ultimately forces the colloid at the bottom (colloid $2$) to cease and for even smaller separation distances $\Delta-2 < 0.3$, the direction of translation reverses. The results found from the method of reflections (solid black lines in Figs.~\ref{Fig_6} and \ref{Fig_7}) can reasonably capture the far field swimming velocities of the particles, however, for smaller separation distances, this method loses its accuracy and more reflected terms should be taken into account for precise results.

Similarly, a catalytically active colloid with uniform reaction at its surface can be propelled due to the presence of another active (or passive) colloid. Fig.~\ref{Fig_7}(a) represents the non-dimensional velocities (scaled with $U_c$) of two catalytically active colloids ($\theta_{cap}^1=\theta_{cap}^2=180^{\circ}$) as a function of their separation distance. When the colloids are far from each other, they do not interact, and since the concentration field is uniform, they do not translate and therefore are stationary. However, when the separation distance becomes smaller, the presence of the other colloid distorts the concentration field around the first one and it eventually yields a propulsion. Likewise, this scenario would take place for a catalytic colloid ($\theta_{cap}^1=180^{\circ}$) and an inert colloid ($\theta_{cap}^2=0^{\circ}$) which can be considered as a cargo, Fig.~\ref{Fig_7}(b). The solid black lines in Fig.~\ref{Fig_7} represent the results found from the method of reflection analysis which suggests that the far field velocity of the catalytic colloids $\sim 1/{\Delta^2}$ and for the inert one $\sim {1}/{\Delta^5}$ (see Appendix~C for details).

\subsubsection{Asymmetric motion of two colloidal motors}
For arbitrary inclination angles of the colloidal particles, they swim in all three directions and they also undergo solid body rotation. In particular, the motions of colloidal motors are influenced not only by hydrodynamic interactions (problem (b), (d) and (e) in Fig.~\ref{Fig_2}) but also via diffusive interaction of the solute concentration generated at the active caps of the colloids with the boundary of the other colloid which modifies the propulsive force (problem (a) in Fig.~\ref{Fig_2}). 
To disentangle these effects better, without loss of generality we focus on situations where the orientation vectors of the particles ${\bf{n}}_p^1$ and ${\bf{n}}_p^2$ are in the same plane $x$ -- $z$ and do not have component in $y$-direction.
\begin{figure} 
\centering
\includegraphics[width=0.99 \textwidth]{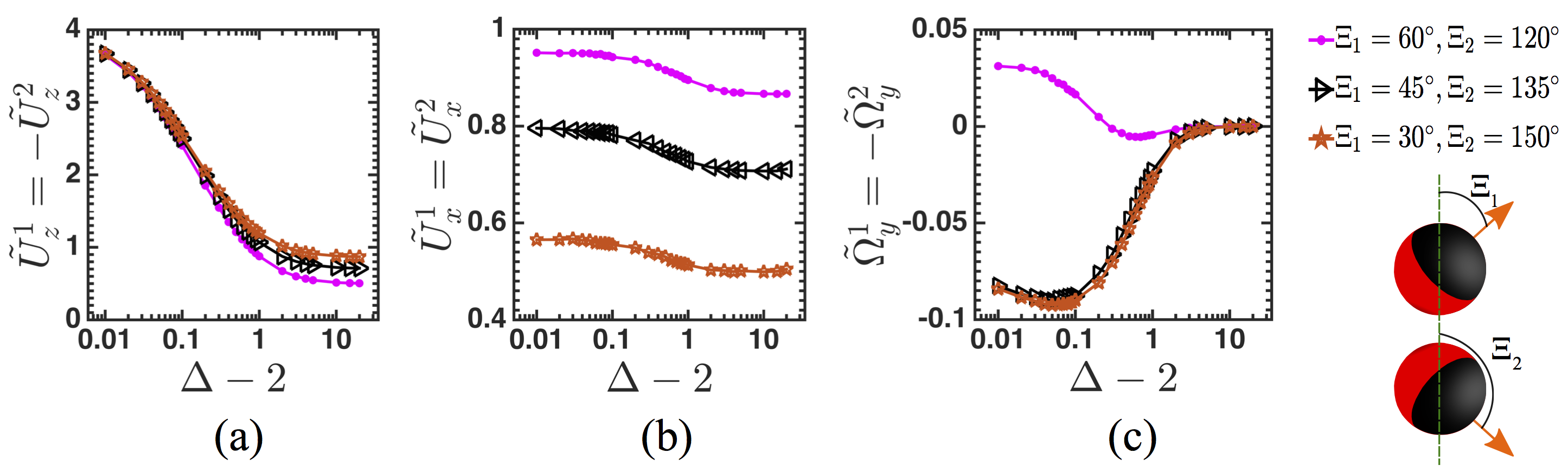}	     	    
 \caption{Non-dimensional translational and angular velocities of two Janus colloidal motors ($\theta_{cap}^1=\theta_{cap}^2=90^{\circ}$) for three distinct relative orientations as a function of non-dimesnional separation distance $\Delta-2$. The translational non-dimensional velocities in the (a) $z$-direction (parallel to the center-to-center line) and (b) $x$-direction (perpendicular to the center-to-center line) are non-dimensionalized with the swimming velocity of a Janus colloid at infinite medium $U_\infty$ and angular velocity (c) around the $y$ axis is non-dimensioqnalized with ${U_\infty}/{a}$.} 
\label{Fig_8}
 \end{figure}
 
As we have observed in the axisymmetric motions, the swimming velocities of colloids depend considerably on their relative orientation, e.g. it does matter in which direction two colloidal motors approach or move away from each other. In order to systematically elucidate this effect, three sets of relative colloid orientations of are considered. The inclination angles of the colloidal particles are chosen to be complimentary, $\Xi_1+\Xi_2=180^{\circ}$ in the first two cases and $\Xi_1-\Xi_2=180^{\circ}$ for the third case. In this way, the magnitudes of swimming translational and angular velocities are identical.
    
In the first scenario, the active areas of the colloidal motors are on one side and adjacent to each other. Fig.~\ref{Fig_8} illustrates the translational and angular velocities of two Janus colloids as a function of separation distance for three relative orientations. The non-dimensional swimming velocity component, which is along the direction of the center-to-center line, approaches the reference value when the colloids are far from each other, i.e. $\tilde{U}^j_z\to \cos\Xi_j$, however, for smaller separation distances, this value increases as the concentration of solute becomes larger in the thin gap for all given inclination angles as shown in Fig.~\ref{Fig_8}(a). The component of non-dimensional swimming velocity in the direction perpendicular to the center-to-center line, $\tilde{U}_x$, behaves similarly; when the motors are far away from each other, it approaches the reference value $\tilde{U}^j_x\to \sin\Xi_j$ and when the separation distance is smaller, the colloids translate faster (Fig.~\ref{Fig_8}(b)). This behavior is predictable based on what was discussed in the axisymmetric motions: For smaller separation distances, the solute concentration is strengthened in the thin gap and brings about stronger propulsions. In addition, the hydrodynamic resistance associated with problem (d) in Fig.~\ref{Fig_2} becomes weaker as the particles translate in the same direction (see \cite{Goldman}) and therefore particles can swim faster. The angular rotation around the $y$ axis, on the other hand, is non-monotonic. To explain this trend, we should first figure out the sign of propulsive torques (torques associated with problem (a) in Fig.~\ref{Fig_2}) on the particles. For complementary relative orientations given in the first scenario, the propulsive torques on the colloids are always equal and opposite, however, depending on the relative orientations, the signs can be positive (clockwise with respect to the negative $y$ axis) or negative (counterclockwise with respect to the negative $y$ axis). To resolve this complex behavior, we have to first elucidate the dependency of the hydrodynamic force associated with problem (d) and (e) in Fig.~\ref{Fig_2} on the separation distance. For two particles translating perpendicular to the line of centers in the same direction under a constant force, the hydrodynamic resistance decreases as the separation distance becomes smaller (see \cite{Goldman}). This reduction in hydrodynamic resistance brings about faster speeds for colloids. The propulsive torque is controlled by the local distribution of the force in the gap which depends on the relative orientations of the colloids and their separation distance. The border line separating the active and passive lobes is the location where the concentration gradient is maximum and, as the the separation distance becomes smaller, the concentration gradient increases at this local region and hence the local force increases. On the other hand, the co-aligned fluid flow generated by two colloids in the gap in this scenario can alleviate the local force in the thin gap and hence the competition between these two brings about the complex trend observed in angular velocity (Fig.~\ref{Fig_8}(c)).
  
 \begin{figure} 
\centering
\includegraphics[width=0.99 \textwidth]{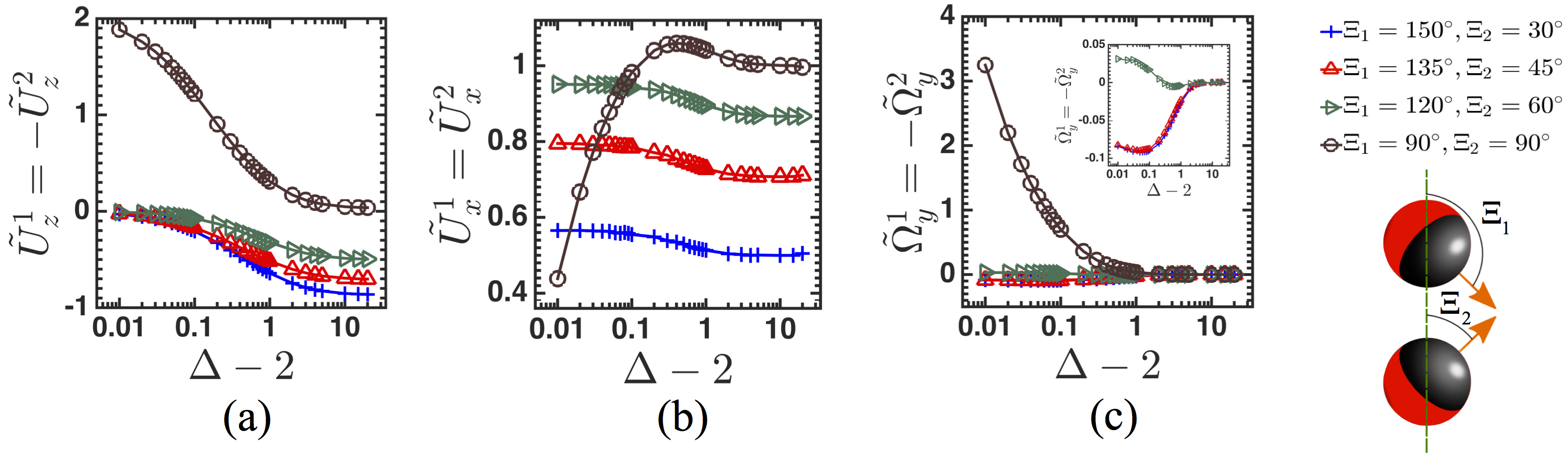}	     	    
 \caption{Non-dimensional translational and angular velocities of two Janus colloidal motors ($\theta_{cap}^1=\theta_{cap}^2=90^{\circ}$) for four distinct relative orientations as a function of non-dimesnional separation distance $\Delta-2$. The translational non-dimensional velocities in (a) $z$-direction (parallel to the center-to-center line) and (b) $x$-direction (perpendicular to the center-to-center line) are non-dimensionalized with the swimming velocity of a Janus colloid at infinite medium $U_\infty$ and angular velocity (c) around the $y$ axis is non-dimensionalized with ${U_\infty}/{a}$. Inset in (c) demonstrate the non-dimensional angular velocity of the relative orientations with small magnitude of rotation.} 
\label{Fig_9}
 \end{figure}
 
In the second scenario, the passive areas of the colloids are next to each other in an inclined angle and they are both translating in the positive $x$-direction (see Fig.~\ref{Fig_9}). Fig.~\ref{Fig_9}(a) shows the $z$ component of the non-dimensional swimming velocity as a function of separation distance for three inclination angles. For all relative inclined orientations, the colloids velocities decrease as the particles swim towards each other and they eventually stops due to the lubrication force in the thin gap. However, for the motion along the positive $x$-direction  where $\Xi_1=\Xi_2=90^{\circ}$, the non-dimensional swimming velocity in the $z$-direction is amplified by the propulsive force due to the presence of excess solute in the gap between the colloids. The swimming velocity component along $x$ and the rotational velocity around $y$ for all relative orientations (except the parallel motion ($\Xi_1=\Xi_2=90^{\circ}$)) behaves exactly as in the previous case, Fig.~\ref{Fig_9}(b) and (c). In the case of parallel motion ($\Xi_1=\Xi_2=90^{\circ}$), the swimming velocity along the $x$-axis increases initially as the separation distance becomes smaller. However, once the separation distance becomes smaller than $\Delta \sim 2.4$, the swimming velocity monotonically decreases (see Fig.~\ref{Fig_9}(b)). The rotational velocity around the $y$ axis in this case is always positive and it monotonically increases as the colloids approach (open circles in Fig.~\ref{Fig_9}(c)). This complex trend is again due to the coupling of the rotational motion around the $y$ axis (positive angular velocity for the colloid on top and negative for the one on the bottom) and positive translational motion in the $x$-direction according to the forces and torques determined from problems (a), (d) and (e) in Fig.~\ref{Fig_2}.     \\ 

 \begin{figure} 
\centering
\includegraphics[width=0.99 \textwidth]{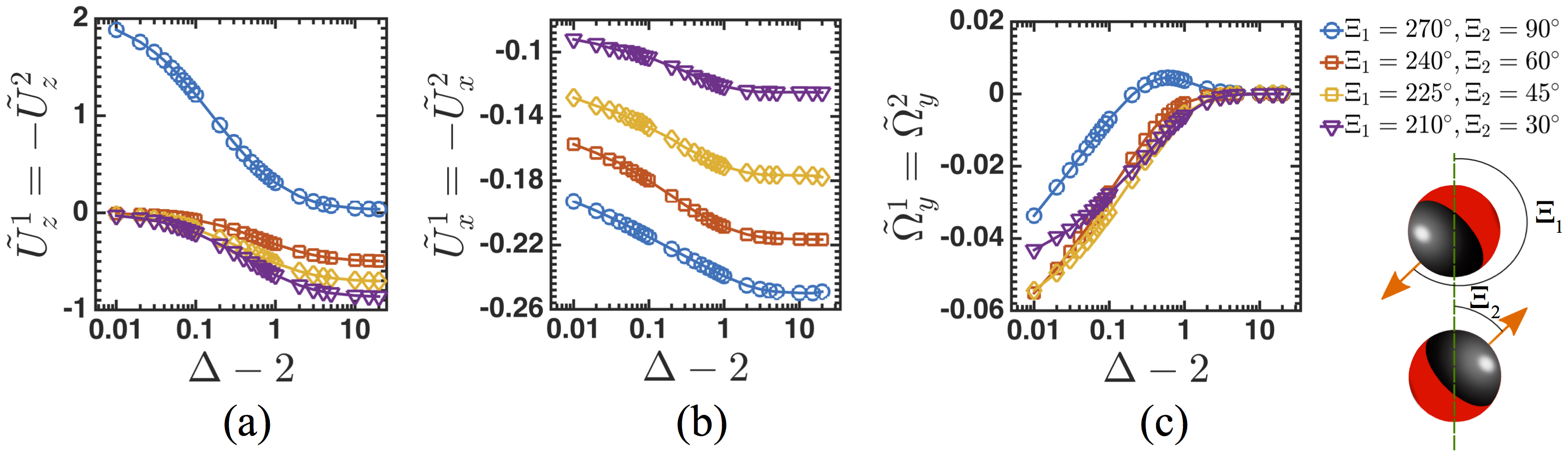}	     	    
 \caption{Non-dimensional translational and angular velocities of two Janus colloidal motors ($\theta_{cap}^1=\theta_{cap}^2=90^{\circ}$) for three distinct relative orientations as a function of non-dimensional separation distance $\Delta-2$. The translational non-dimensional velocities in (a) $z$-direction (parallel to the center-to-center line) and (b) $x$-direction (perpendicular to the center-to-center line) are non-dimensionalized with the swimming velocity of a Janus colloid at infinite medium $U_\infty$ and angular velocity (c) around the $y$ axis is non-dimensionalized with ${U_\infty}/{a}$.} 
\label{Fig_10}
 \end{figure}

In the third scenario, the active caps face opposite to each other in all the given orientations, Fig.~\ref{Fig_10}. The $z$ component of the swimming velocity is identical to the second scenario, Fig.~\ref{Fig_10}(a). However, the swimming velocity in the $x$-direction and angular velocity behave differently. Fig.~\ref{Fig_10}(b) demonstrates the non-dimensional swimming velocity in the $x$-direction as a function of separation distance. The speed of colloids for all orientations decreases as they move towards each other. To realize the trend in this scenario, we allude again to the hydrodynamic forces associated with problems (d) and (e) in Fig.~\ref{Fig_2}. For two passive colloidal particles which are translating in opposite directions under a constant force \cite{ONeill}, the hydrodynamic torques on the two colloids are equal in sign and magnitude and hence the particles rotate in the same direction, e.g. if two passive colloids on the top and bottom of the $x-y$ plane translate in negative and positive $x$-directions, respectively, the hydrodynamic torque forces both colloids to rotate clockwise with respect to the negative $y$-axis. Moreover, the hydrodynamic resistance to the motion of two passive colloids translating in opposite directions along the $x$-axis increases as the separation distance decreases and therefore results in smaller translational velocity. Fig.~\ref{Fig_10}(b) demonstrates the non-dimensional swimming velocity in the $x$-direction as a function of separation distance; the speed of colloids for all of the orientations decreases as they move towards each other since the hydrodynamic resistance scales as $\sim \log(1/({\Delta-2}))$ \cite{Kim}. The colloids also rotate faster as they approach each other (see Fig.~\ref{Fig_10}(c)). For the parallel motion ($\Xi_1=270^{\circ}$ and $\Xi_2 = 90^{\circ}$), the rotational velocity is initially positive and then it undergoes a transition to the negative values. The initial positive values are due to the fact that the propulsive torque in this case is always positive, i.e. the torques associated with problem (a) in Fig.~\ref{Fig_2} always tend to force the colloids to rotate in a way that they bring their active sections close to each other and this effect is dominant in the far field once $\Xi_1=270^{\circ}$ and $\Xi_2 = 90^{\circ}$. However, once the colloids approach each other, the hydrodynamic torque associated with problem (d) in Fig.~\ref{Fig_2} offsets this positive torque and ultimately compels the colloids to rotate with negative angular velocities.
 
 \subsection{Pair trajectories}
To investigate the pair trajectories of colloidal motors, we assume that, similar to previous section, the initial orientation vectors of the motors are both in the $x$ -- $z$ plane and consequently the colloids stay in the same plane for long time since it was shown that there are no rotations around the $y$-axis and hence the colloids never exit the $x$ -- $z$ plane at long time. Generally, for two colloidal particles approaching each other, there are two ultimate possible scenarios: either they (i) interact and ``escape'' from each other or (ii) they persistently approach and come into contact to assemble.  We refer to  ``assembly'' when the non-dimensional separation distance is less than $\Delta < 2.01$. Here, complementary angles for relative orientations are considered, i.e. the initial inclination angles with respect to the line of centers are either $\Xi_1+\Xi_2=180^{\circ}$ or $\Xi_1-\Xi_2=180^{\circ}$; the former indicates the conditions where the $x$ component of swimming velocities of both motors are in the positive direction while the latter includes situations where the $x$ component of swimming velocities are in  opposite directions.
\begin{figure} 
\centering
\includegraphics[width=0.75 \textwidth]{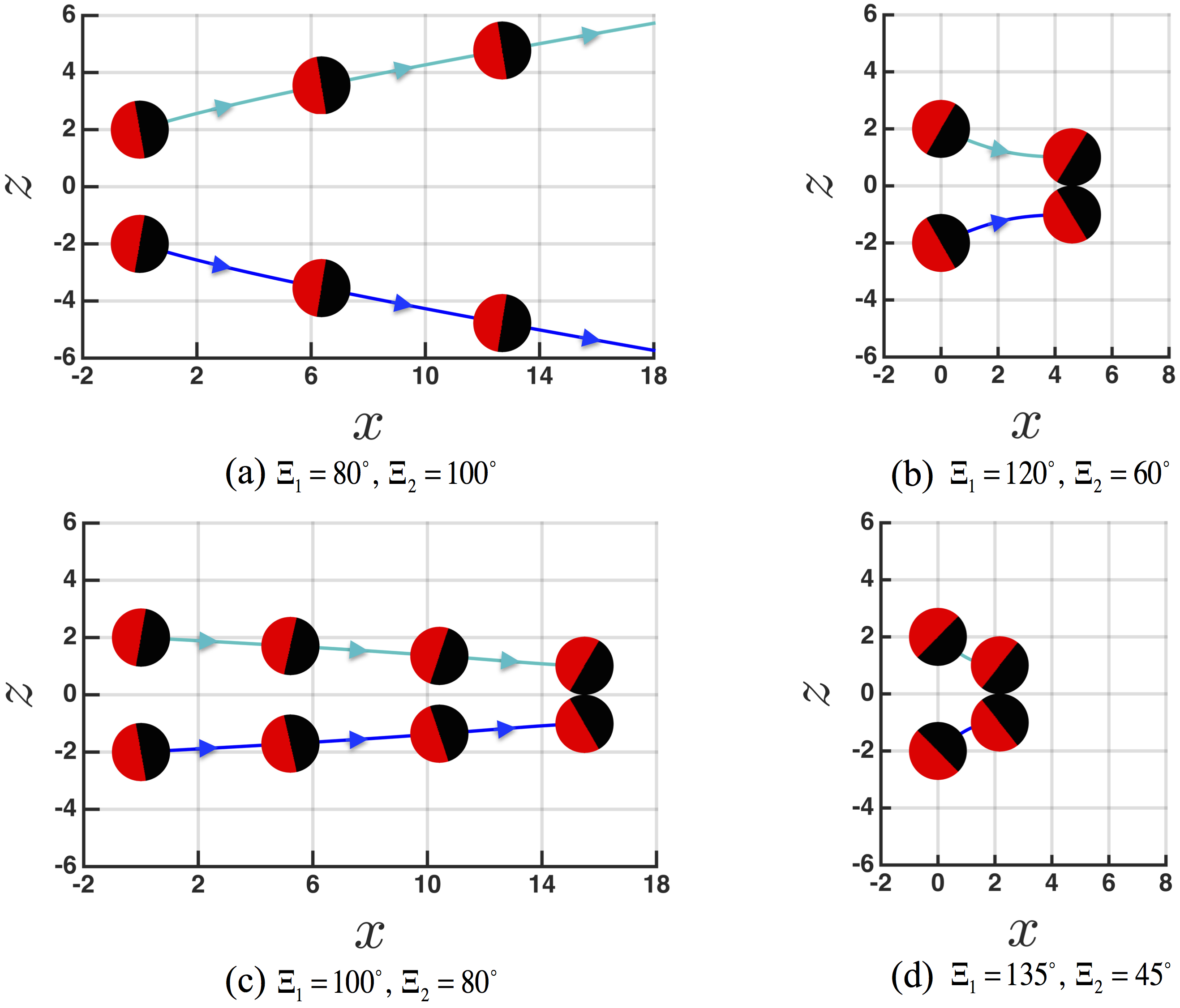}	     	    
 \caption{Pair trajectories of Janus motors for various initial inclination angles where $\Xi_1+\Xi_2=180^{\circ}$.} 
\label{Fig_11}
 \end{figure}
 For two swimming colloids, the direction of rotation can be altered depending on two parameters: (i) relative orientations and (ii) colloids active coverages. 
 
 The relative orientation indicates whether the active sections of the colloidal motors are adjacent or not. If the initial orientation of the motors is set in a fashion similar to the first scenario (see the schematic in Fig.~\ref{Fig_8}), the build-up of the excess concentration of solute in the gap between the particles gives rise to a situation where particles persistently move away from each other until they fail to interact and each one of them swims independently (see Fig.~\ref{Fig_11}(a)).

If the initial relative orientation of the motors is given similar to the second scenario as shown in Fig.~\ref{Fig_9}, the colloids approach and rotate at the same time until they assemble. For two Janus motors, the direction of rotation at most relative orientations are in away that enhance the assembly process of the particles, i.e. the particles rotate in opposite directions in a way to bring their passive sections aligned and closer. The pair trajectory of two Janus colloids with various initial relative orientations and separation distances of $\Delta =4$ is shown in Fig.~\ref{Fig_11}(b), (c) and (d). For initial inclination angles $\Xi_1=100^{\circ}$ (Fig.~\ref{Fig_11}(b)) and $\Xi_1=120^{\circ}$ (Fig.~\ref{Fig_11}(c)), the particle on top (bottom) rotates clockwise (counterclockwise) with respect to the negative $y$-direction and this in turn enhances the assembly of the particles. For $\Xi_1=135^{\circ}$ (Fig.~\ref{Fig_11}(d)), the directions of rotations of the colloids reversed, however, the magnitude of rotations are too small to allow particles to escape. \\
\begin{figure} 
\centering
\includegraphics[width=0.75 \textwidth]{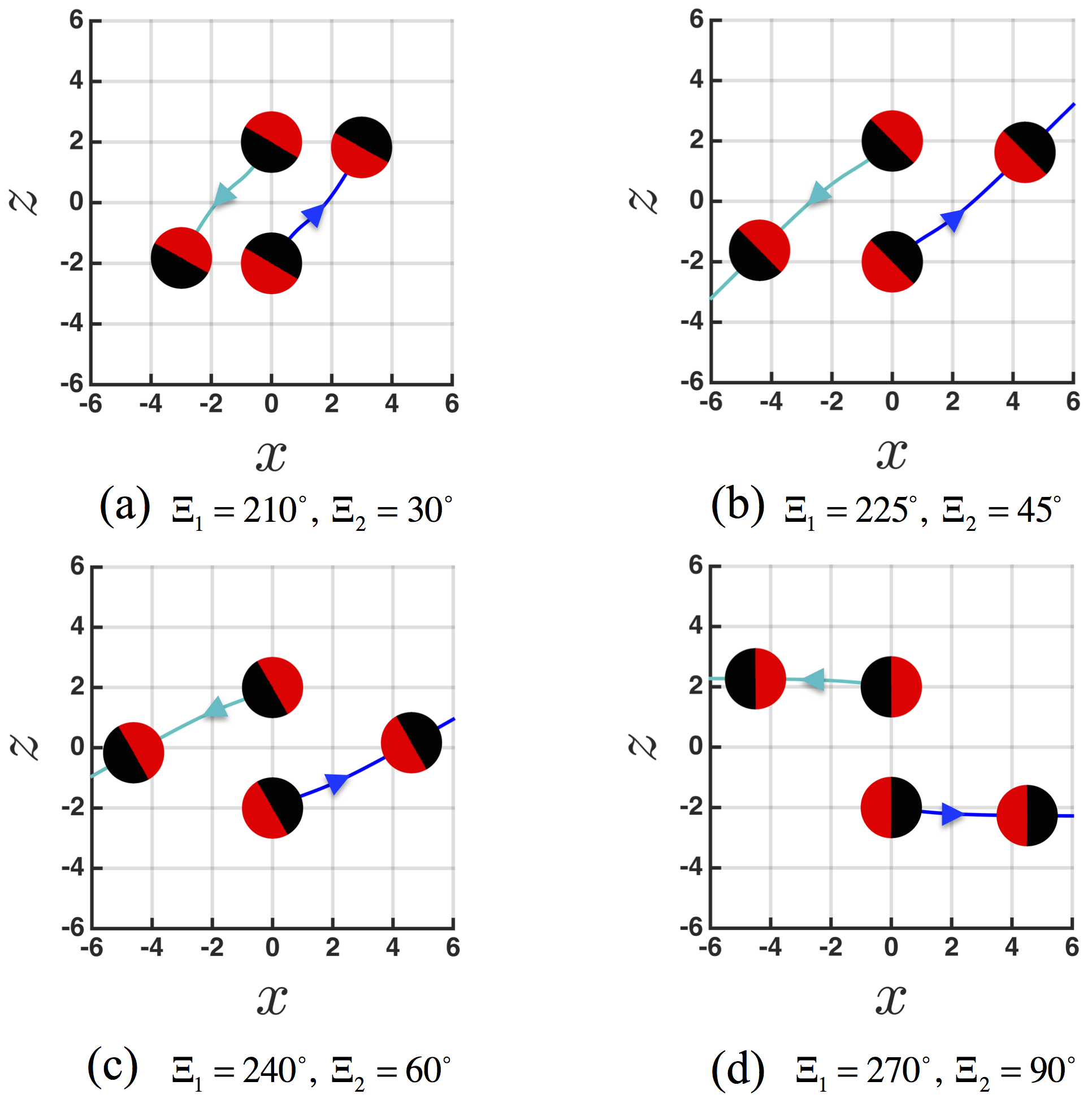}	     	    
 \caption{Pair trajectories of Janus motors for various relative initial inclination angels where $\Xi_1-\Xi_2=180^{\circ}$.} 
\label{Fig_12}
 \end{figure}
 
In the third scenario, as shown in Fig.~\ref{Fig_10}, the passive sections of colloids are adjacent. The typical pair trajectory of two Janus motors for various initial inclination angles $\Xi_1$ are illustrated in Fig.~\ref{Fig_12}. Here, the colloids always rotate in the same direction around the $y$-axis. The particle tend to face their active sections and for all initial inclination angles $\Xi_1>194^{\circ}$ at $\Delta =4$, the colloids pass each other and escape. For smaller initial inclination angle $\Xi_1\le194^{\circ}$, the particles block each other's paths and hence they assemble.
 \begin{figure} 
\centering
\includegraphics[width=0.99 \textwidth]{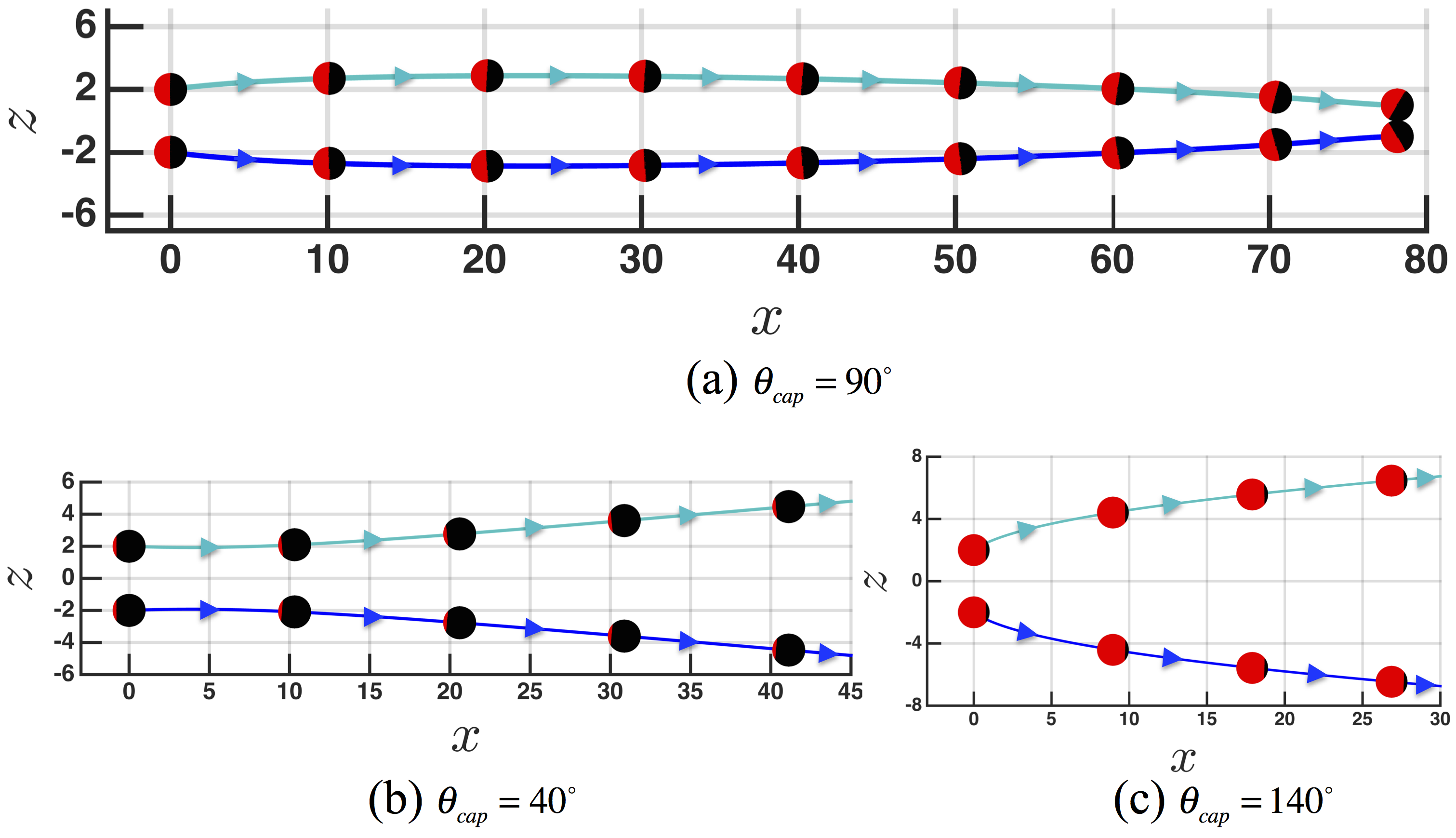}	     	    
 \caption{Pair trajectories of catalytically active motors for a fixed relative initial inclination angels of $\Xi_1=\Xi_2=90^{\circ}$ and (a) $\theta_{cap}=40^{\circ}$, (b) $\theta_{cap}=140^{\circ}$ and (c) $\theta_{cap}=90^{\circ}$.} 
\label{Fig_13}
 \end{figure}
 
In all three scenarios, the active coverages of the colloidal motors play an important role on the pair trajectories. To resolve this fact more clearly, in Fig.~\ref{Fig_13} we demonstrate the pair trajectories of active colloidal motors for a fixed relative orientation of $\Xi_1=\Xi_2=90^{\circ}$ and three different cap angles of $\theta^{j}_{cap}=40^{\circ}$, $\theta^{j}_{cap}=90^{\circ}$ and $\theta^{j}_{cap}=140^{\circ}$. In Fig.~\ref{Fig_13}(a), the pair trajectory of two Janus motors $\theta^{j}_{cap}=90^{\circ}$ is shown; the particles initially move away from each other because the swimming velocity along the $z$-direction is positive (see Fig.~\ref{Fig_9}(a)). However, since the angular velocity around the $y$ axis is positive (negative) for the particle on top (bottom) (see Fig.~\ref{Fig_9}(c)), the colloids rotate to bring their passive sections facing to the gap and approach to assemble. For $\theta^{j}_{cap}=40^{\circ}$, the direction of rotations of the particles around the $y$ axis are reversed. This type of pair trajectory is a typical behavior of two swimming bacteria whose dynamics are usually studied in the context of squirmer model \cite{Lighthill,Blake,Pedley}. For larger coverages, the directions of rotation of the particles around the $y$-axis are also reversed. However, the presence of excess solute concentrated in the gap repels the particles to the distant regions where the particles do not interact and therefore swim independently (see Fig.~\ref{Fig_13} (c)). Moreover, for this initial separation distance $\Delta=4$ and inclination angles of $\Xi_1=\Xi_2=90^{\circ}$, there is a critical cap angle (coverage) $\theta_{cap}^1=\theta_{cap}^2=63^{\circ}$ in which the propulsive torque precisely cancels the hydrodynamic torques of problems (d) and (e) in figure.~\ref{Fig_2} and hence the particles do not rotate. However, this condition is unstable because the swimming velocity component in the $z$-direction is nonzero and hence the particles move away from each other.

\begin{figure} 
\centering
\includegraphics[width=0.85 \textwidth]{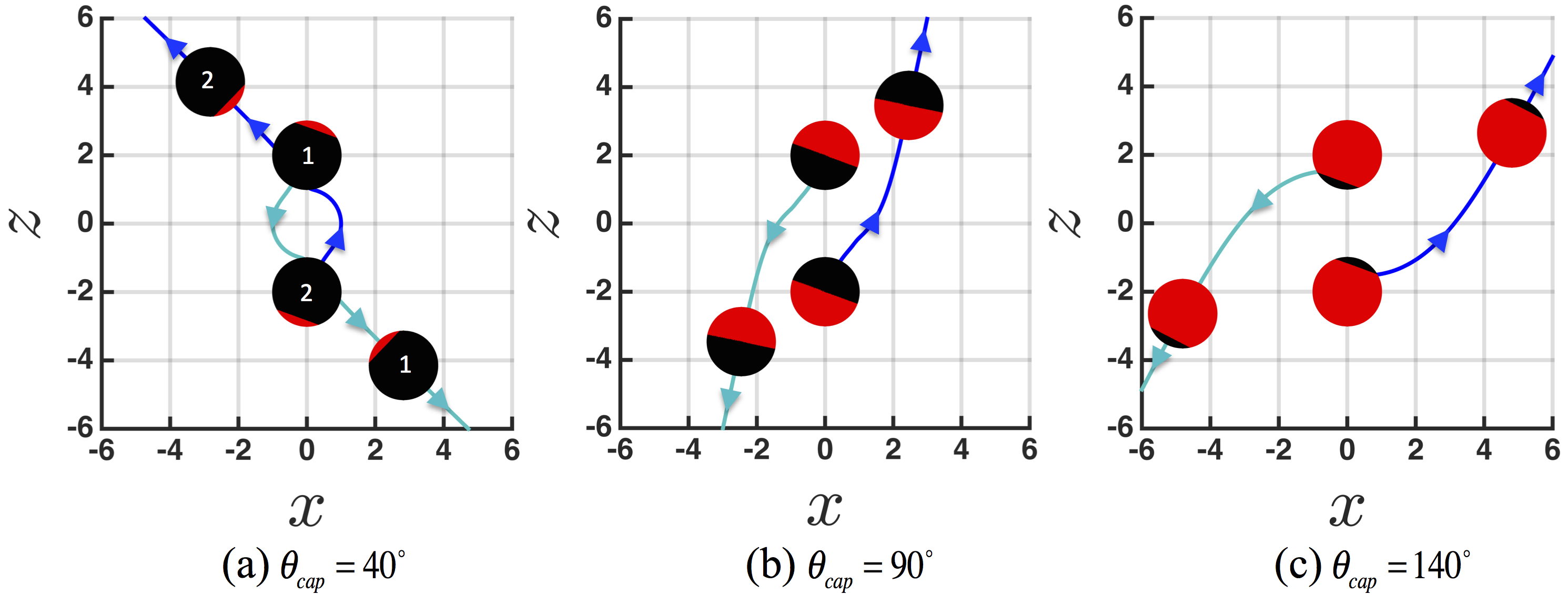}	     	    
 \caption{Pair trajectories of catalytically active motors for a fixed relative initial inclination angle $\Xi_1= 200^{\circ},~\Xi_2=20^{\circ}$ and (a) $\theta_{cap}=40^{\circ}$, (b) $\theta_{cap}=140^{\circ}$ and (c) $\theta_{cap}=90^{\circ}$. The numbers on (a) are for clarity on particles trajectories.} 
\label{Fig_14}
 \end{figure}

Fig.~\ref{Fig_14} shows the pair trajectories that correspond to the third scenario (see Fig.~\ref{Fig_10}) for three different cap angles. In all cases, the shape of the trajectories are almost similar. However, the direction of rotations are not similar for all coverages. For $\theta^{j}_{cap}=40^{\circ}$ and $\theta^{j}_{cap}=90^{\circ}$ the particles rotate with negative angular velocities since the particles can reach smaller separation distances where the positive propulsive torques (torque associated with problem (a) in Fig.~\ref{Fig_2}) are small compared to the negative hydrodynamic torques generated by problem (d) in Fig.~\ref{Fig_2}. For high coverage $\theta^{j}_{cap}=140^{\circ}$ on the other hand, the particles are so far that the colloids have positive angular velocities. Moreover, as the particle coverage becomes smaller in size, the particles stay in the sticky lubrication zone much longer and this leads to a more curved pair trajectory (see Fig.~\ref{Fig_14}(a)).

\begin{figure} 
\centering
\includegraphics[width=0.95 \textwidth]{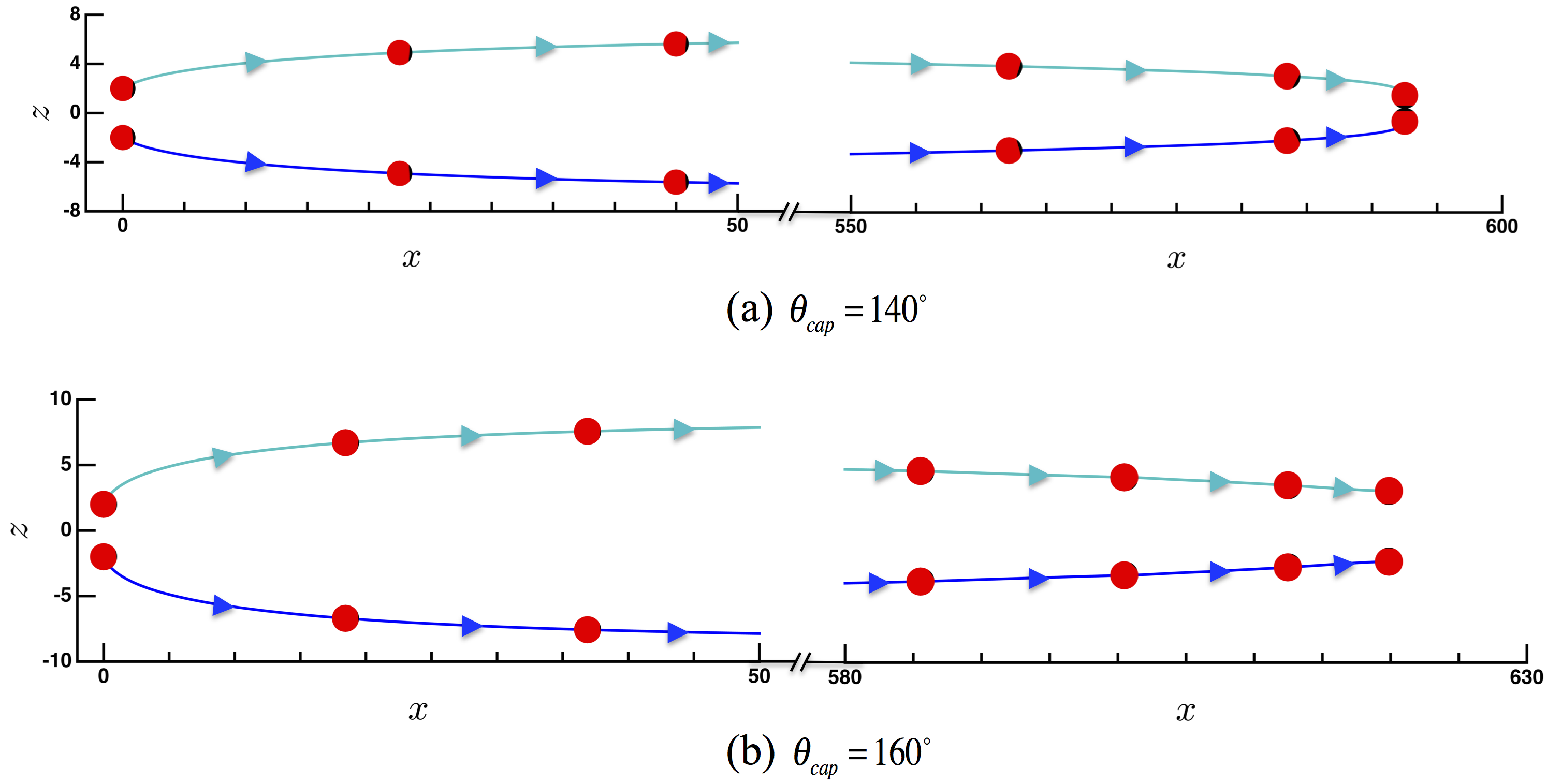}	     	    
 \caption{Pair trajectories of catalytically active motors for a fixed initial inclination angles of $\Xi_1= 97^{\circ},~\Xi_2=83^{\circ}$ and (a) $\theta_{cap}=140^{\circ}$ (b) $\theta_{cap}=160^{\circ}$. } 
\label{Fig_16}
 \end{figure}

For particles with very high coverages, the particles initially move away to distant regions where they attain positive (particle on top) and negative (particle on bottom) rotations and translate until they reach an axisymmetric orientation, i.e. $180^{\circ}$ for the one on top and  $0^\circ$ for the one on bottom, and then approach to assemble (see Fig.~\ref{Fig_16}(a)). Furthermore, the analysis indicates beyond a critical coverage of $\theta_{cap}=152^{\circ}$ the particles can never come into contact. However, there is one interesting feature in the pair trajectories where the motors become completely stationary (see \ref{Fig_16}(b)). At the stationary point, the concentration field around each one of the particles is uniform and hence the propulsive forces/torques are identically zero. Thus they come to a complete stop and then remain stationary indefinitely. This type of trajectory has also been observed for catalytic motors with high coverages in the vicinity of a solid boundary \cite{Uspal,Wall_1}. 
 
 \section{Conclusions}\label{Conc}
A combined analytical-numerical solution has been developed to study the pair interaction of catalytically active colloids due to self-diffusiophoresis. These colloids act as ``motors'' consuming fuel from the immediate environment and create a local concentration gradients of interactive solutes which in turn drive their autonomous motion. Since the driving force for particle propulsion is solute concentration gradient, we solved for both hydrodynamics and mass transfer equations around two colloidal particles utilizing a continuum approach for diffusiophoresis \cite{a89} in the limits of $Da\ll 1$ and $Pe\ll1$. The mass transfer problem was formulated and solved exactly in bispherical coordinates. For hydrodynamics, we took advantage of the Reynolds Reciprocal Theorem \cite{hapbren83} to write the propulsive forces and torques on colloids as quadratures.

The results presented here show that the dynamics and trajectories of two catalytically active colloids are much richer and more complex than other self-propelled particles that solely interact via hydrodynamics, i.e. colloidal squirmers. For catalytic particles, the presence of a second particle distorts the solute concentration around the first particle and therefore modifies its propulsive force and torque. Thus, employing purely hydrodynamic models with a constant force or source-dipoles fails to capture the actual dynamics of catalytically driven active colloids. As evident in these systems, the local concentration gradient of solutes can push (or pull) the colloids from the domains with high solute concentration resulting in depleted (or aggregated) zones. The hydrodynamic interaction is the second effect which influences the dynamics and particle trajectories.  In conclusion, the dynamics at the pair level are controlled by the active coverage of catalytic caps $\theta^{j}_{cap}$ and the relative orientation of the colloids $\Xi_j,~(j=1$ and $2)$.

When the active areas of the colloids are aligned, the motions are axisymmetric and particles do not rotate. For situations where active surfaces are adjacent, the solute concentration rises in the gap between the colloids, which causes them to experience an increase in propulsive force. In contrast, when the passive surfaces of the particles are adjacent, the propulsive forces do not changes significantly while particles are approaching each other and therefore their swimming velocities are reduced by lubrication forces.

 \begin{figure} 
\centering
\includegraphics[width=0.85 \textwidth]{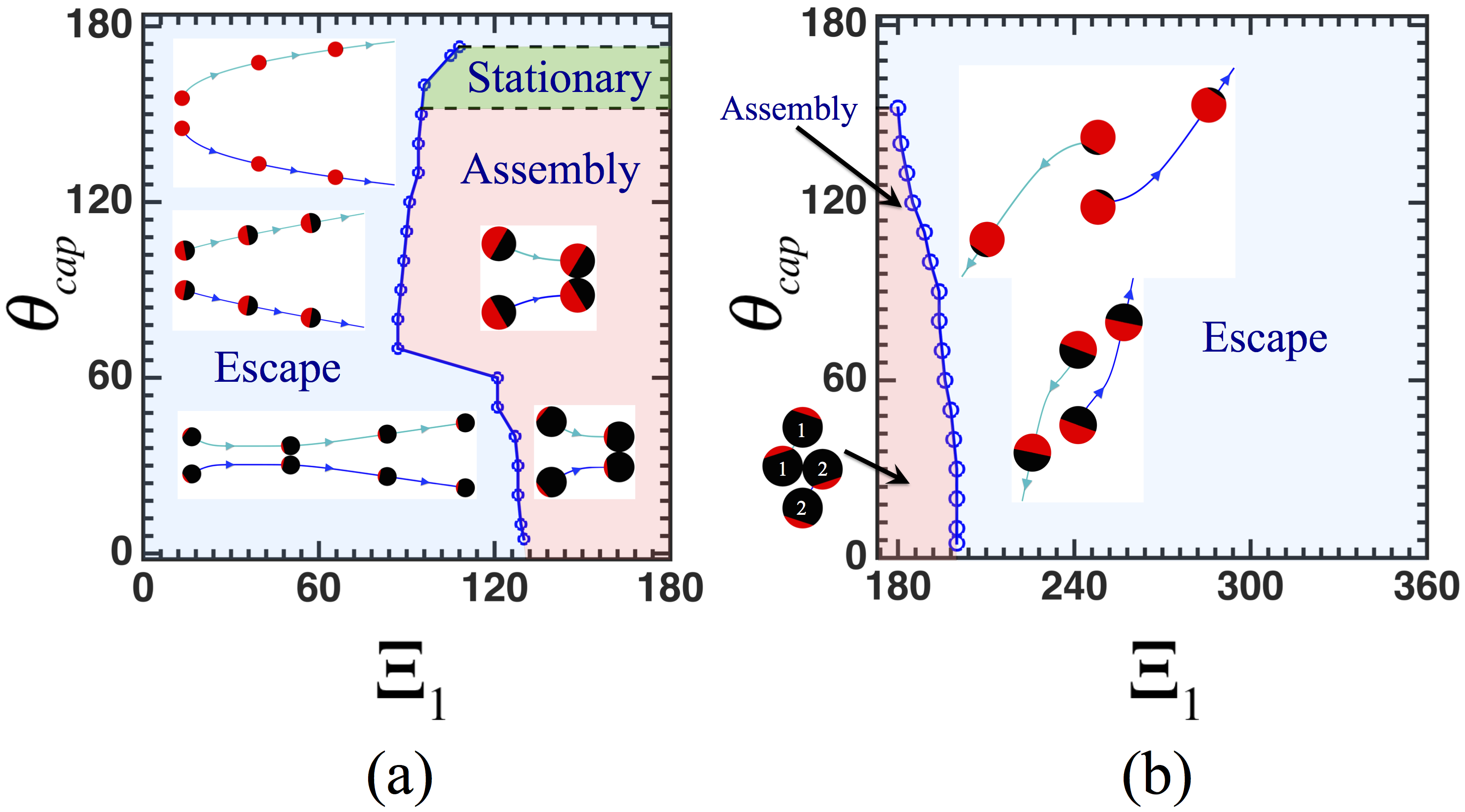}	     	    
 \caption{Phase diagrams of pair trajectories of catalytically active colloids with identical coverage $\theta_{cap}$ and initial separation distance of $\Delta=4$ for complementary angles of (a) $\Xi_1+\Xi_2=180^{\circ}$ and (b) $\Xi_1-\Xi_2=180^{\circ}$. } 
\label{Fig_15}
 \end{figure}
 
When the active areas of the colloids are not aligned completely, the motions are asymmetric and the colloids rotate and translate at the same time. For situations where the orientation vectors of the colloidal motors are in the $x$ -- $z$ plane (see Fig.~\ref{Fig_1}), there are two possible scenarios for the particles: colloids can either swim towards each other and come into contact (assembly) or they can move away from each other (escape). The direction of colloids rotations determine which one of the two scenarios takes place. Based on the predictions of this model, a phase diagram was sketched in Fig.~\ref{Fig_15} for long time behavior of two non-Brownian particles with similar coverage $\theta_{cap}$ at an initial separation distance of $\Delta=4$ and complementary relative orientations of $\Xi_1+\Xi_2=180^{\circ}$ (Fig.~\ref{Fig_15}(a)) and $\Xi_1-\Xi_2=180^{\circ}$ (Fig.~\ref{Fig_15}(b)). For small coverages, there are greater angles for initial relative orientation of the particles to come into contact while for larger coverage the condition of assembly occurs under more stringent conditions and beyond a critical coverage of $\theta_{cap}=152^{\circ}$ the particles never come into contact. We believe this phase diagram has important implications on the elucidation of the clustering and phase separation of catalytically self-propelled particles from a mechanical perspective and could be useful for experimentalists. Finally, the pair level analysis we undertook in the present study can be a basis for pursuing much more complex systems in understanding the role of hydrodynamics in active matter. In particular, a suspension of catalytically active particles has a unique feature in which the active particles can chemically signal each other and be pulled (or pushed) to the certain regions in the system.
However, a detailed and quantitative analysis of the collective behavior of catalytically driven active particles requires taking into account two formidable challenges: (i) the many body hydrodynamic and phoretic interactions and (ii)  Brownian motion. This can be accomplished by employing Stokesian dynamics simulations \cite{SDS} which we will address in a future communication.

\section*{Acknowledgements}

U.M.C.-F. is supported by National Science Foundation grants CBET -- 1055284 and EPS -- 1010674.\\

\appendix
  \section{Recursive relations for the concentration field around two colloids}\label{apa}
In this section, we provide the recursive relationships found by applying the boundary conditions on particle surfaces for the concentration field,
\begin{align}
-&\left[ {n\sinh (n - 0.5){\beta _j}} \right]{{\tilde A}_{n - 1,0}} + \left[ {\sinh {\beta _j}\cosh (n + 0.5){\beta _j} + } \right.\nonumber\\
&\left. {(2n + 1)\cosh {\beta _j}\sinh (n + 0.5){\beta _j}} \right]{{\tilde A}_{n,0}} - \left[ {(n + 1)\sinh (n + 1.5){\beta _j}} \right]{{\tilde A}_{n + 1,0}} - \nonumber\\
&\left[ {n\cosh (n - 0.5){\beta _j}} \right]{{\tilde B}_{n - 1,0}} + \left[ {\sinh {\beta _j}\sinh (n + 0.5){\beta _j} + } \right.\nonumber\\
&\left. {(2n + 1)\cosh {\beta _j}\cosh (n + 0.5){\beta _j}} \right]{{\tilde B}_{n,0}} - \left[ {(n + 1)\cosh (n + 1.5){\beta _j}} \right]{{\tilde B}_{n + 1,0}}\nonumber\\
 &= \Psi _{n,0}^j,\label{rec-m-0}
\end{align}  
\begin{align}
 -&\left[ {(n - m)\sinh (n - 0.5){\beta _j}} \right]{{\tilde A}_{n - 1,m}} + \left[ {\sinh {\beta _j}\cosh (n + 0.5){\beta _j} + } \right.\nonumber\\
&\left. {(2n + 1)\cosh {\beta _j}\sinh (n + 0.5){\beta _j}} \right]{{\tilde A}_{n,m}} - \left[ {(n + m + 1)\sinh (n + 1.5){\beta _j}} \right]{{\tilde A}_{n + 1,m}} - \nonumber\\
&\left[ {(n - m)\cosh (n - 0.5){\beta _j}} \right]{{\tilde B}_{n - 1,m}} + \left[ {\sinh {\beta _j}\sinh (n + 0.5){\beta _j} + } \right.\nonumber\\
&\left. {(2n + 1)\cosh {\beta _j}\cosh (n + 0.5){\beta _j}} \right]{{\tilde B}_{n,m}} - \left[ {(n + m + 1)\cosh (n + 1.5){\beta _j}} \right]{{\tilde B}_{n + 1,m}}\nonumber\\
&= \Psi _{n,m}^j,\label{rec-m-n-1}
\end{align}
\begin{align}
 - &\left[ {(n - m)\sinh (n - 0.5){\beta _j}} \right]{{\tilde C}_{n - 1,m}} + \left[ {\sinh {\beta _j}\cosh (n + 0.5){\beta _j} + } \right.\nonumber\\
&\left. {(2n + 1)\cosh {\beta _j}\sinh (n + 0.5){\beta _j}} \right]{{\tilde C}_{n,m}} - \left[ {(n + m + 1)\sinh (n + 1.5){\beta _j}} \right]{{\tilde C}_{n + 1,m}} -\nonumber \\
&\left[ {(n - m)\cosh (n - 0.5){\beta _j}} \right]{{\tilde D}_{n - 1,m}} + \left[ {\sinh {\beta _j}\sinh (n + 0.5){\beta _j} + } \right.\nonumber \\
&\left. {(2n + 1)\cosh {\beta _j}\cosh (n + 0.5){\beta _j}} \right]{{\tilde D}_{n,m}} - \left[ {(n + m + 1)\cosh (n + 1.5){\beta _j}} \right]{{\tilde D}_{n + 1,m}}\nonumber\\
&= {\rm{Z}}_{n,m}^j,\label{rec-m-n-2}
\end{align}
where we have
\begin{align}
&\Psi _{n,0}^j = \frac{{2n + 1}}{{4\pi }}\int_0^{2\pi } {d\phi \int_0^\pi  {\frac{{2c {\vartheta _j}{u_j}}}{{\sqrt {\cosh {\beta _j} - \cos \alpha } }}\sin \alpha~ P_n^0(\cos \alpha )d\alpha } } ,\\
&\Psi _{n,m}^j = \frac{{(2n + 1)(n - m)!}}{{2\pi (n + m)!}}\int_0^{2\pi } {d\phi \int_0^\pi  {\frac{{2c {\vartheta _j}{u_j}\cos m\phi }}{{\sqrt {\cosh {\beta _j} - \cos \alpha } }}\sin \alpha ~P_n^m(\cos \alpha )d\alpha } } ,\\
&{\rm Z}_{n,m}^j = \frac{{(2n + 1)(n - m)!}}{{2\pi (n + m)!}}\int_0^{2\pi } {d\phi \int_0^\pi  {\frac{{2c {\vartheta _j}{u_j}\sin m\phi }}{{\sqrt {\cosh {\beta _j} - \cos \alpha } }}\sin \alpha~ P_n^m(\cos \alpha )d\alpha } } .
\end{align}
Note that each one of the the Eqs.~\ref{rec-m-0} -- \ref{rec-m-n-2} are two equations for $j=1$ and $2$.
 \section{Stokes flow solution of two identical spherical colloids}
In this section we provide the detailed solution of Stokes flow around two identical colloids. For arbitrary motion of colloids, the motion can be described as translational motions parallel and perpendicular along the center-to-center lines and rotational motion around the third axis. To calculate the hydrodynamic forces and torques on colloids and utilizing Reynolds Reciprocal Theorem (RRT), we recapitulate the hydrodynamic solutions in this section.\\
The Stokes flow solution of the velocity field in cylindrical coordinates $(\rho,\phi ,z)$ in the form of eigensolutions of bispherical coordinates is given by
\begin{align}
p(\alpha ,\beta ,\phi ) = \frac{{\sqrt {\cosh \beta  - \cos \alpha } }}{c}\sum\limits_{n = 1}^\infty  {\left( {{A_n}\sinh (n + \frac{1}{2})\beta  + {B_n}\cosh (n + \frac{1}{2})\beta } \right)P_n^1(\cos \alpha )\cos \phi } ,\label{aux1} 
\end{align}
\begin{equation}
\begin{split}
&{v_\rho } = \frac{{\sin \alpha }}{{2\sqrt {\cosh \beta  - \cos \alpha } }}\sum\limits_{n = 1}^\infty  {\left( {{A_n}\sinh (n + \frac{1}{2})\beta  + {B_n}\cosh (n + \frac{1}{2})\beta } \right)\;P_n^1(\cos \alpha )\;\cos \phi } \\
&+ \frac{{\sqrt {\cosh \beta  - \cos \alpha } }}{2}\;\left[ {\sum\limits_{n = 2}^\infty  {\left( {{E_n}\sinh (n + \frac{1}{2})\beta  + {F_n}\cosh (n + \frac{1}{2})\beta } \right)P_n^2(\cos \alpha )\cos \phi \;} } \right.\\
&\left. { + \sum\limits_{n = 0}^\infty  {\left( {{G_n}\sinh (n + \frac{1}{2})\beta  + {H_n}\cosh (n + \frac{1}{2})\beta } \right)\;P_n^0(\cos \alpha )\;\cos \phi } } \right],\label{vel1}
\end{split}
\end{equation}
\begin{equation}
\begin{split}
&{v_\phi } = \frac{{\sqrt {\cosh \beta  - \cos \alpha } }}{2}\left[ {\sum\limits_{n = 2}^\infty  {\left( {{E_n}\sinh (n + \frac{1}{2})\beta  + {F_n}\cosh (n + \frac{1}{2})\beta } \right)\;P_n^2(\cos \alpha )\;\sin \phi } } \right.\\
&\left. { - \sum\limits_{n = 0}^\infty  \; \left( {{G_n}\sinh (n + \frac{1}{2})\beta  + {H_n}\cosh (n + \frac{1}{2})\beta } \right)P_n^0(\cos \alpha )\;\sin \phi } \right],\label{vel2}
\end{split}
\end{equation}
\begin{equation}
\begin{split}
&{v_z} = \frac{{\sinh \beta }}{{2\sqrt {\cosh \beta  - \cos \alpha } }}\sum\limits_{n = 1}^\infty  {\left( {{A_n}\sinh (n + \frac{1}{2})\beta  + {B_n}\cosh (n + \frac{1}{2})\beta } \right)P_n^1(\cos \alpha )\;} \cos \phi \;\\
&+ \sqrt {\cosh \beta  - \cos \alpha } \;\sum\limits_{n = 1}^\infty  {\left( {{C_n}\sinh (n + \frac{1}{2})\beta  + {D_n}\cosh (n + \frac{1}{2})\beta } \right)P_n^1(\cos \alpha )}\cos \phi,\label{vel3}
\end{split}
\end{equation} 
where 8 sets of unknown coefficients, namely ${A_{n} }$, ${B_{n} }$,..., ${H_{n} }$, have to be obtained from satisfying the boundary conditions at the surface of two particles   and also the equation of continuity.\\
To study the translational and rotational motions perpendicular to the line of centers and around the third axis, we have to find the following four solutions.
\subsection{Slow translation of two identical spheres normal to their line of centers and in the same direction}\label{vnormal}
The slow translation of two non-rotating identical spheres where the two particles are moving in the same direction and with the same velocity $U_{\bot}^{1}$ perpendicular to their line of centers is considered here. Boundary conditions at the surface of two particles are equivalent and are given by:
\begin{align}
 &\left. {v_r } \right|_{\beta  = \beta _1 }  = U_{\bot}^{1}\;\cos \phi,  \nonumber\\ 
 &\left. {v_\phi  } \right|_{\beta  = \beta _1 }  =  - U_{\bot}^{1}\;\sin \phi,  \label{bcmaj}\\ 
 &\left. {v_z } \right|_{\beta  = \beta _1 }  = 0. \nonumber 
\end{align}
According to the boundary condition, one can easily realize that $\gamma_0=0$ . In the above equations, $\beta  = \beta _1$ represents the surface of a spherical particle and if $\beta _1$ is positive it is located above the $x-y$ plane. It can be easily shown that the only non zero coefficients are ${B_{n} }$, ${C_{n} }$, ${F_{n} }$ and ${H_{n} }$ and, using boundary conditions given in Eqs.~\ref{bcmaj}, this yields the following recursive relations among these four coefficients
\begin{align}
&- \frac{1}{{2(2n - 1)}}\cosh (n - \frac{1}{2}){\beta _1}\;{B_{n - 1}} + \frac{1}{{2(2n + 3)}}\cosh (n + \frac{3}{2}){\beta _1}\;{B_{n + 1}}\nonumber\\
&- \frac{{(n - 2)}}{{(2n - 1)}}\cosh (n - \frac{1}{2}){\beta _1}\;{F_{n - 1}} + \cosh {\beta _1}\cosh (n + \frac{1}{2}){\beta _1}\;{F_n}\nonumber\\
&- \frac{{(n + 3)}}{{(2n + 3)}}\;\cosh (n + \frac{3}{2}){\beta _1}\;{F_{n + 1}} = 0,\label{maj1} 
\end{align}

\begin{align}
&\frac{{(n - 1)\;n}}{{2(2n - 1)}}\cosh (n - \frac{1}{2}){\beta _1}\;{B_{n - 1}} - \frac{{(n + 1)(n + 2)}}{{2(2n + 3)}}\cosh (n + \frac{3}{2}){\beta _1}\;{B_{n + 1}}\nonumber\\
&- \frac{n}{{(2n - 1)}}\cosh (n - \frac{1}{2}){\beta _1}{H_{n - 1}} + \cosh {\beta _1}\cosh (n + \frac{1}{2}){\beta _1}\;{H_n}\nonumber\\
&- \frac{{(n + 1)}}{{(2n + 3)}}\cosh (n + \frac{3}{2}){\beta _1}\;{H_{n + 1}} = 2\sqrt 2 \;\left[ {\cosh {\beta _1}\;{e^{ - (n + \frac{1}{2}){\beta _1}}}} \right.\nonumber \\
&\left. { - (\frac{n}{{2n - 1}}){e^{ - (n - \frac{1}{2}){\beta _1}}} - (\frac{{n + 1}}{{2n + 3}}){e^{ - (n + \frac{3}{2}){\beta _1}}}} \right],\label{maj2}
\end{align}

\begin{align}
 &\frac{{\sinh \beta _1 }}{2}\;\cosh (n + \frac{1}{2})\beta _1 \;B_n  - \frac{{(n - 1)}}{{(2n - 1)}}\sinh (n - \frac{1}{2})\beta _1 \;C_{n - 1} \nonumber \\ 
 & + \cosh \beta _1 \sinh (n + \frac{1}{2})\beta _1 \;C_{n}  - \frac{{(n + 2)}}{{(2n + 3)}}\sinh (n + \frac{3}{2})\beta _1 \;C_{n + 1}  = 0. \label{maj3}
\end{align}
Finally the velocity components should satisfy the equation of continuity,
\begin{align}
&\frac{{ - 1}}{2}(n - 1)\;B_{n - 1}  + \frac{5}{2}\;B_n  + \frac{1}{2}(n + 2)\;B_{n + 1}  - (n - 1)\;C_{n - 1}\nonumber  \\ 
 &+ (2n + 1)C_n  - (n + 2)C_{n + 1}  - \frac{1}{2}(n - 2)(n - 1)F_{n - 1}  + (n + 2)(n - 1)F_n \nonumber \\ 
 &- \frac{1}{2}(n + 2)(n + 3)F_{n + 1}  + \frac{1}{2}H_{n - 1}  - H_n  + \frac{1}{2}H_{n + 1}  = 0. \label{maj4}
\end{align}
Eq.~\ref{maj1} is valid for $n \ge 2$, Eq.~\ref{maj2} for $n \ge 0$ and Eqs.~\ref{maj3} -- \ref{maj4}  for $n \ge 1$. Four sets of unknown coefficients are obtained from simultaneous solution of the above equations and, as the particles separation distances becomes smaller, more terms in the series should be kept to reach an accurate result.   
The hydrodynamic forces on the colloids are similar in magnitude and sign, however, the torques are equal in magnitude but opposite in signs:
\begin{equation}
\mathcal{F}^{j,1}_{\bot}   =  - 2\sqrt {2\;} \pi \;\sinh \beta _1 \;\mu \;U_{\bot}^1\,a\;\sum\limits_{n = 0}^\infty  {H_n },
\end{equation}
\begin{equation}
{\mathcal T}_ \otimes ^{j,1} =  \pm 2\sqrt {2\;} \pi \;{\left( {\sinh {\beta _1}} \right)^2}\;\mu \;U_ \bot ^1\,{a^2}\;\sum\limits_{n = 0}^\infty  {\left( {2n + 1 - \coth {\beta _1}} \right){H_n}}, 
\end{equation}
where $j =1$ or $2$ refers to the colloid at positive and negative $z$, respectively. 
\subsection{Slow rotation of two identical spheres about an axis perpendicular to their line of centers and in opposite directions}\label{majrot}
In this section, we solve equations of motion and continuity for two identical non translating spheres that are rotating around the third axis that passes through their center and normal to their line of centers with equal angular velocity $\Omega_{\otimes }^2$ but in the opposite directions. The relevant boundary conditions at the surface of first sphere are given by
\begin{align}
&{\left. {{v_r}} \right|_{\beta  = {\beta _1}}} = \Omega _ \otimes ^2\left( {z - {\Delta  \mathord{\left/
 {\vphantom {\Delta  2}} \right.
 \kern-\nulldelimiterspace} 2}} \right)\cos \phi,\nonumber  \\ 
&{\left. {{v_\phi }} \right|_{\beta  = {\beta _1}}} =  - \Omega _ \otimes ^2\left( {z - {\Delta  \mathord{\left/
 {\vphantom {\Delta  2}} \right.
 \kern-\nulldelimiterspace} 2}} \right)\sin \phi,\label{bcmaj2}  \\ 
&\left. {v_z } \right|_{\beta  = \beta _1 }  = -\Omega_{\otimes }^2 \;r\cos \phi,\nonumber
 \end{align}
and the boundary conditions at the surface of the second sphere are
\begin{align}
&{\left. {{v_r}} \right|_{\beta  =  - {\beta _1}}} =  - \Omega _ \otimes ^2\left( {z + {\Delta  \mathord{\left/
 {\vphantom {\Delta  2}} \right.
 \kern-\nulldelimiterspace} 2}} \right)\cos \phi, \nonumber \\ 
&{\left. {{v_\phi }} \right|_{\beta  =  - {\beta _1}}} = \Omega _ \otimes ^2\left( {z + {\Delta  \mathord{\left/
 {\vphantom {\Delta  2}} \right.
 \kern-\nulldelimiterspace} 2}} \right)\sin \phi,\label{bcmaj22}  \\ 
&\left. {v_z } \right|_{\beta  = -\beta _1 }  = \Omega_{\otimes }^2\; r\cos \phi. \nonumber 
\end{align}
Applying conditions \ref{bcmaj2} - \ref{bcmaj22} in cylindrical components of equations of motion, Eqs.~\ref{vel1} -- \ref{vel3}, it turns out that the only non-vanishing coefficents are ${B_{n} }$, ${C_{n} }$, ${F_{n} }$ and ${H_{n} }$. Four relationships are needed to obtain this set of four unknown coefficients and it can be readily shown that two following recursive relationships along with Eqs.~\ref{maj1} and \ref{maj4} solves the problem:
\begin{align}
&\frac{{(n - 1)\;n}}{{2(2n - 1)}}\cosh (n - \frac{1}{2}){\beta _1}\;{B_{n - 1}} - \frac{{(n + 1)(n + 2)}}{{2(2n + 3)}}\cosh (n + \frac{3}{2}){\beta _1}\;{B_{n + 1}}\nonumber\\
&- \frac{n}{{(2n - 1)}}\cosh (n - \frac{1}{2}){\beta _1}{H_{n - 1}} + \cosh {\beta _1}\cosh (n + \frac{1}{2}){\beta _1}\;{H_n}\nonumber\\
&- \frac{{(n + 1)}}{{(2n + 3)}}\cosh (n + \frac{3}{2}){\beta _1}\;{H_{n + 1}} = 2\sqrt 2 \;c\left[ { - \frac{1}{{\sinh {\beta _1}}}\;{e^{ - (n + \frac{1}{2}){\beta _1}}}} \right.\\
&\left. { + \coth {\beta _1}(\frac{n}{{2n - 1}}){e^{ - (n - \frac{1}{2}){\beta _1}}} + \coth {\beta _1}(\frac{{n + 1}}{{2n + 3}}){e^{ - (n + \frac{3}{2}){\beta _1}}}} \right],\;\;(n \ge 0),\label{maj5}
\end{align}

\begin{align}
 &\frac{{\sinh \beta _1 }}{2}\;\cosh (n + \frac{1}{2})\beta _1 \;B_n  - \frac{{(n - 1)}}{{(2n - 1)}}\sinh (n - \frac{1}{2})\beta _1 \;C_{n - 1} \nonumber \\ 
  &+ \cosh \beta _1 \sinh (n + \frac{1}{2})\beta _1 \;C_{n}  - \frac{{(n + 2)}}{{(2n + 3)}}\sinh (n + \frac{3}{2})\beta _1 \;C_{n + 1}  = \nonumber \\ 
  &- \sqrt 2 \;c \;\left[{(\frac{{ - 1}}{{2n - 1}})e^{ - (n - \frac{1}{2})\beta _1 }  + (\frac{1}{{2n + 3}})e^{ - (n + \frac{3}{2})\beta _1 } )}\right],\;\;(n \ge 1). \label{maj6}
\end{align}
The hydrodynamic force on the colloids are similar in magnitude and sign however, the torques are equal in magnitude but opposite in signs:
\begin{equation}
\mathcal{F}^{j,2}_{\bot}=  2\sqrt {2\;} \pi \;\sinh \beta _1 \;\mu \;\Omega_{\otimes}^2 \,a^2 \;\sum\limits_{n = 0}^\infty  {H_n }, 
\end{equation}
\begin{equation}
{\mathcal T}_ \otimes ^{j,2} =  \mp 2\sqrt {2\;} \pi \;{\left( {\sinh {\beta _1}} \right)^2}\;\mu \;\Omega _ \otimes ^2\,{a^3}\;\sum\limits_{n = 0}^\infty  {\left( {2n + 1 - \coth {\beta _1}} \right){H_n}},
\end{equation}
where $j =1$ or $2$ refers to the colloid at positive and negative $z$, respectively.
\subsection{Slow translation of two identical spheres with similar velocities in magnitude but opposite directions along the direction perpendicular to their line of centers}
In this problem, one sphere is translating without rotation with velocity $U_{\bot}^3$ towards the positive direction normal to the line of centers while the other identical sphere moves with the velocity $-U_{\bot}^3$ towards the negative direction or vice versa. 
The boundary conditions at each point on the surface of a sphere whose center is located in $z>0$ can be expressed as
\begin{align}
&\left. {v_r } \right|_{\beta  = \beta _1 }  = U_{\bot}^{3}\;\cos \phi, \nonumber \\ 
&\left. {v_\phi  } \right|_{\beta  = \beta _1 }  =  - U_{\bot}^{3}\;\sin \phi,  \\ 
&\left. {v_z } \right|_{\beta  = \beta _1 }  = 0.\nonumber
\end{align}
For the sphere whose center is located at negative values of $z$, the related boundary conditions for this non-rotating sphere is given by
\begin{align}
 &\left. {v_r } \right|_{\beta  = -\beta _1 }  = -U_{\bot}^{3}\;\cos \phi,  \nonumber\\ 
 &\left. {v_\phi  } \right|_{\beta  = -\beta _1 }  = U_{\bot}^{3}\;\sin \phi,  \\ 
 &\left. {v_z } \right|_{\beta  =-\beta _1 }  = 0. \nonumber
\end{align}
It can be immediately shown that coefficients ${B_{n} }$, ${C_{n} }$, ${F_{n} }$ and ${H_{n} }$ in the general solution of the    Stokes flow, \ref{vel1}-\ref{vel3}, are zero. In order to find the remaining coefficients, the following sets of relations (which can be determined by satisfying appropriate boundary conditions at the particle surfaces as well as the equation of continuity) have to be solved simultaneously:
\begin{align}
 &- \frac{1}{{2(2n - 1)}}\sinh (n - \frac{1}{2}){\beta _1}\;{A_{n - 1}} + \frac{1}{{2(2n + 3)}}\sinh (n + \frac{3}{2}){\beta _1}\;{A_{n + 1}}\nonumber\\
 &- \frac{{(n - 2)}}{{(2n - 1)}}\sinh (n - \frac{1}{2}){\beta _1}\;{E_{n - 1}} + \cosh {\beta _1}\sinh (n + \frac{1}{2}){\beta _1}\;{E_n}\nonumber\\
 &- \frac{{(n + 3)}}{{(2n + 3)}}\;\sinh (n + \frac{3}{2}){\beta _1}\;{E_{n + 1}} = 0,\;\;\;(n \ge 2), \label{on1}
\end{align}

\begin{align}
&\frac{{(n - 1)\;n}}{{2(2n - 1)}}\sinh (n - \frac{1}{2}){\beta _1}\;{A_{n - 1}} - \frac{{(n + 1)(n + 2)}}{{2(2n + 3)}}\sinh (n + \frac{3}{2}){\beta _1}\;{A_{n + 1}}\nonumber\\
&- \frac{n}{{(2n - 1)}}\sinh (n - \frac{1}{2}){\beta _1}{G_{n - 1}} + \cosh {\beta _1}\sinh (n + \frac{1}{2}){\beta _1}\;{G_n}\nonumber\\
&- \frac{{(n + 1)}}{{(2n + 3)}}\sinh (n + \frac{3}{2}){\beta _1}\;{G_{n + 1}} = 2\sqrt 2 \;\left[ {\cosh {\beta _1}\;{e^{ - (n + \frac{1}{2}){\beta _1}}}} \right.\\
&\left. { - (\frac{n}{{2n - 1}}){e^{ - (n - \frac{1}{2}){\beta _1}}} - (\frac{{n + 1}}{{2n + 3}}){e^{ - (n + \frac{3}{2}){\beta _1}}}} \right],\;\;\;(n \ge 0), \label{on2}
\end{align}

\begin{align}
&\frac{{\sinh \beta _1 }}{2}\;\sinh (n + \frac{1}{2})\beta _1 \;A_n  - \frac{{(n - 1)}}{{(2n - 1)}}\cosh (n - \frac{1}{2})\beta _1 \;D_{n - 1} \nonumber \\ 
 & + \cosh \beta _1 \cosh (n + \frac{1}{2})\beta _1 \;D_{n}  - \frac{{(n + 2)}}{{(2n + 3)}}\cosh (n + \frac{3}{2})\beta _1 \;D_{n + 1}  = 0,\;\;(n \ge 1),\label{on3}
\end{align}

\begin{align}
& \frac{{ - 1}}{2}(n - 1)\;A_{n - 1}  + \frac{5}{2}\;A_n  + \frac{1}{2}(n + 2)\;A_{n + 1}  - (n - 1)\;D_{n - 1} \nonumber \\ 
&  + (2n + 1)D_n  - (n + 2)D_{n + 1}  - \frac{1}{2}(n - 2)(n - 1)E_{n - 1}  + (n + 2)(n - 1)E_n  \nonumber\\ 
&  - \frac{1}{2}(n + 2)(n + 3)E_{n + 1}  + \frac{1}{2}G_{n - 1}  - G_n  + \frac{1}{2}G_{n + 1}  = 0,\;\;(n \ge 1). \label{on4}
\end{align}
Two particles experience force in the $x$-direction with the same magnitude but opposite sign. The hydrodynamic torques are in the $y$-direction and are equal. They can be given by the following infinite series relations for the particle in the positive half space
\begin{equation}
\mathcal{F}^{j,3}_{\bot}   =  \mp 2\sqrt {2\;} \pi \;\sinh \beta _1 \;\mu \;U\,a\;\sum\limits_{n = 0}^\infty  {G_n }, 
\end{equation}
\begin{equation}
{\mathcal T}_ \otimes ^{j,3} = 2\sqrt {2\;} \pi \;{\left( {\sinh {\beta _1}} \right)^2}\;\mu \;U\,{a^2}\;\sum\limits_{n = 0}^\infty  {\left( {2n + 1 - \coth {\beta _1}} \right){G_n}},
\end{equation}
where $j =1$ or $2$ refers to the colloid at positive and negative $z$, respectively.
\subsection{Slow rotation of two equal spheres about an axis perpendicular to their line of centers and in a same direction}
In this problem, two non-translating spheres are rotating with the same angular velocities, $\Omega_{\otimes}^{4}$, around their diameter which is parallel to the $y$-axis. Cylindrical components of the velocity field are satisfied at the particle surface of the sphere which is located above the $x-y$ plane if it is written as
\begin{align}
&{\left. {{v_r}} \right|_{\beta  = {\beta _1}}} = \Omega _ \otimes ^4\left( {z - {\Delta  \mathord{\left/
 {\vphantom {\Delta  2}} \right.
 \kern-\nulldelimiterspace} 2}} \right)\cos \phi, \nonumber \\ 
&{\left. {{v_\phi }} \right|_{\beta  = {\beta _1}}} =  - \Omega _ \otimes ^4\left( {z - {\Delta  \mathord{\left/
 {\vphantom {\Delta  2}} \right.
 \kern-\nulldelimiterspace} 2}} \right)\sin \phi, \nonumber \\ 
&\left. {v_z } \right|_{\beta  = \beta _1 }  =  - \Omega_{\otimes}^{4} \;r\cos \phi,  
\end{align}
and for the second sphere as 
\begin{align}
 &{\left. {{v_r}} \right|_{\beta  =  - {\beta _1}}} = \Omega _ \otimes ^4\;\left( {z + {\Delta  \mathord{\left/
 {\vphantom {\Delta  2}} \right.
 \kern-\nulldelimiterspace} 2}} \right)\cos \phi, \nonumber  \\ 
 &{\left. {{v_\phi }} \right|_{\beta  =  - {\beta _1}}} =  - \Omega _ \otimes ^4\;\left( {z + {\Delta  \mathord{\left/
 {\vphantom {\Delta  2}} \right.
 \kern-\nulldelimiterspace} 2}} \right)\sin \phi,\nonumber \\ 
 &\left. {v_z } \right|_{\beta  = -\beta _1 }  =  - \Omega_{\otimes}^{4} \;r\cos \phi.  
\end{align}
Upon satisfying boundary conditions at the surface of two particles, the two following recursive relations among four non-zero coefficients (${A_{n} }$, ${D_{n} }$, ${E_{n} }$ and ${G_{n} }$) are obtained: 

\begin{align}
&\frac{{(n - 1)\;n}}{{2(2n - 1)}}\sinh (n - \frac{1}{2}){\beta _1}\;{A_{n - 1}} - \frac{{(n + 1)(n + 2)}}{{2(2n + 3)}}\sinh (n + \frac{3}{2}){\beta _1}\;{A_{n + 1}}\nonumber\\
&- \frac{n}{{(2n - 1)}}\sinh (n - \frac{1}{2}){\beta _1}{G_{n - 1}} + \cosh {\beta _1}\sinh (n + \frac{1}{2}){\beta _1}\;{G_n}\nonumber\\
&- \frac{{(n + 1)}}{{(2n + 3)}}\sinh (n + \frac{3}{2}){\beta _1}\;{G_{n + 1}} = 2\sqrt 2 \;c\left[ { - \frac{1}{{\sinh {\beta _1}}}\;{e^{ - (n + \frac{1}{2}){\beta _1}}}} \right.\\
&\left. { + \coth {\beta _1}(\frac{n}{{2n - 1}}){e^{ - (n - \frac{1}{2}){\beta _1}}} + \coth {\beta _1}(\frac{{n + 1}}{{2n + 3}}){e^{ - (n + \frac{3}{2}){\beta _1}}}} \right],\;\;(n \ge 0), 
\end{align}

\begin{align}
& \frac{{\sinh \beta _1 }}{2}\;\sinh (n + \frac{1}{2})\beta _1 \;A_n  - \frac{{(n - 1)}}{{(2n - 1)}}\sinh (n - \frac{1}{2})\beta _1 \;D_{n - 1} \nonumber \\ 
 & + \cosh \beta _1 \sinh (n + \frac{1}{2})\beta _1 \;D_{n}  - \frac{{(n + 2)}}{{(2n + 3)}}\sinh (n + \frac{3}{2})\beta _1 \;D_{n + 1}    \nonumber\\ 
&  =- \sqrt 2 \;c\;\left[ {(\frac{{ - 1}}{{2n - 1}}){e^{ - (n - \frac{1}{2}){\beta _1}}} + (\frac{1}{{2n + 3}}){e^{ - (n + \frac{3}{2}){\beta _1}}})} \right],\;\;(n \ge 1).
\end{align}
Eqs.~\ref{on1} and \ref{on4}, which is an equation of continuity, stay the same. The hydrodynamic force and torque on these two particles are equal and can be found from following infinite series relations:
\begin{equation}
\mathcal{F}^{j,4}_{\bot}   =  \pm 2\sqrt {2\;} \pi \;\sinh \beta _1 \;\mu \;\Omega_{\otimes}^{4} \,a^2 \;\sum\limits_{n = 0}^\infty  {G_n }, 
\end{equation}
\begin{equation}
{\mathcal T}_ \otimes ^{j,4} =  - 2\sqrt {2\;} \pi \;{\left( {\sinh {\beta _1}} \right)^2}\;\mu \;\Omega _ \otimes ^4\,{a^3}\;\sum\limits_{n = 0}^\infty  {\left( {2n + 1 - \coth {\beta _1}} \right){G_n}},
\end{equation}
where $j =1$ or $2$ refers to the colloid at positive and negative $z$, respectively.
\subsection{Slow rotations of two identical spheres about an axis along their line of centers}\label{Tnormal}
This problem has been solved by Jeffrey by stream function solution because of the axisymmetric nature of problem. We address this problem by directly solving the Stokes equation in cylindrical coordinates. The general form of the solution in this case can be written as,
\begin{align}
{v_\phi } = \sqrt {\cosh \beta  - \cos \alpha } \;\sum\limits_{n = 1}^\infty  {\left( {{G_{n,0}}\sinh (n + \frac{1}{2})\beta  + {H_{n,0}}\cosh (n + \frac{1}{2})\beta } \right)P_n^1(\cos \alpha )}.
\end{align}
Here, we consider two different scenarios:\\

 (I) Two particles are rotating in the same direction with a boundary condition that can be written as
\begin{eqnarray}
{\left. {{v_\phi }} \right|_{\beta  = {\beta _j}}} = \frac{{c\sin \alpha }}{{\cosh {\beta _j} - \cos \alpha }}\Omega _\parallel ^1,
\end{eqnarray}
from which one can instantly obtain:
\begin{eqnarray}
&{G_{n,0}} = 0,
\end{eqnarray}
\begin{align}
&{H_{n,0}} = \sqrt 2 c\frac{{{e^{ - (n + \frac{3}{2}){\beta _1}}} - {{\mathop{\rm e}\nolimits} ^{ - (n - \frac{1}{2}){\beta _1}}}}}{{\sinh {\beta _1}\cosh (n + \frac{1}{2}){\beta _1}}}.
\end{align}
In this case, the hydrodynamic torque on the particle can be written as, 
\begin{eqnarray}
{\mathcal T}_\parallel ^1 = 4\sqrt {2\;} \pi \;{\left( {\sinh {\beta _1}} \right)^2}\mu \; \Omega _\parallel ^1\;{a^3}\sum\limits_{n = 1}^\infty  {n\left( {n + 1} \right){H_{n,0}}}.
\end{eqnarray}

(II) Two particles are rotating in opposite directions with a boundary condition can be written as
\begin{eqnarray}
{\left. {{v_\phi }} \right|_{\beta  = {\beta _j}}} = \pm \frac{{c\sin \alpha }}{{\cosh {\beta _j} - \cos \alpha }}\Omega _\parallel ^2,
\end{eqnarray}
from which one can instantly obtain:
\begin{align}
&{G_{n,0}} = \sqrt 2 c\frac{{{e^{ - (n + \frac{3}{2}){\beta _1}}} - {e^{ - (n - \frac{1}{2}){\beta _1}}}}}{{\sinh {\beta _1}\sinh (n + \frac{1}{2}){\beta _1}}},
\end{align}
\begin{align}
&{H_{n,0}} = 0,
\end{align}
and the hydrodynamic torque on the particle can be written as, 
\begin{eqnarray}
{\mathcal T}_\parallel ^2 =  \pm 4\sqrt {2\;} \pi \;{\left( {\sinh {\beta _1}} \right)^2}\mu\; \Omega _\parallel ^2\;{a^3}\sum\limits_{n = 1}^\infty  {n\left( {n + 1} \right){G_{n,0}}}.
\end{eqnarray}
Consequently, any rotations around the $z$ axis can be written as a linear combination of the these two problems. 
\section{Method of reflections}
\subsection{Two partially coated catalytic colloids}
\begin{figure} 
\centering
\includegraphics[width=0.35 \textwidth]{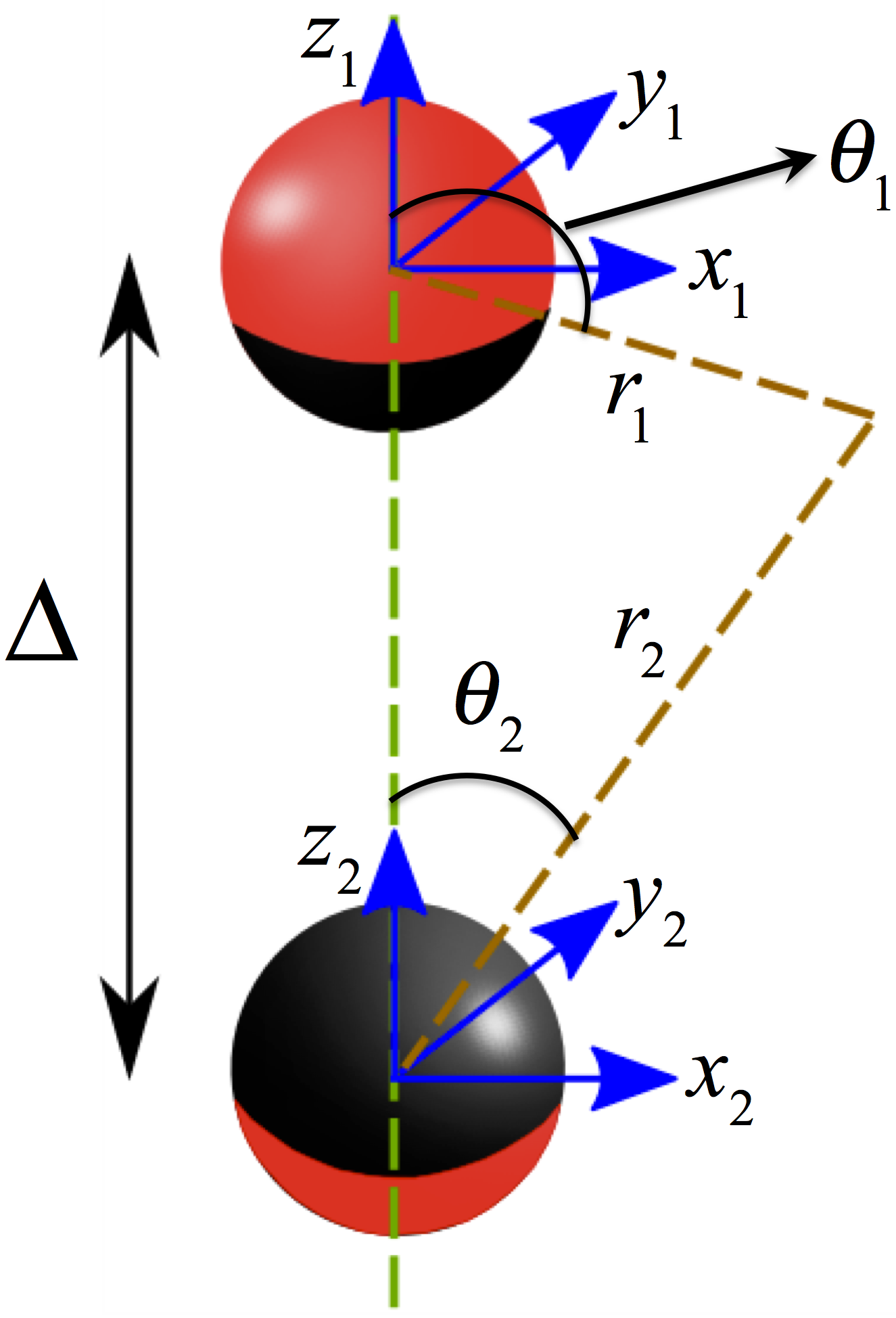}	     	    
 \caption{Schematic representation of two aligned catalytically active colloids.} 
\label{Fig_App}
 \end{figure}
 Suppose we have a pair of catalytically active motors ($1$ and $2$ stands for the one on top and bottom, respectively; see Fig.~\ref{Fig_App} for detail) with constant flux production of solute. The non-dimensional concentration field ($\sim N_2a\ D$) around the first motor in the absence of the second motor can be found according to the following:
 \begin{eqnarray}
{n_2}({r_2},{\theta _2}) = \sum\limits_{n = 0}^\infty  {\frac{{{d_n}}}{{r_2^{n + 1}}}{P_n}(\cos {\theta _2})},
 \end{eqnarray}
 where ${P_n}$ is the Legendre polynomials and $d_n$ are constants be determined via the following,
 \begin{align}
{d_n} = \frac{{(2n + 1)}}{2(n+1)}\int_0^\pi  {{u_2(\theta_2)}~\sin\theta_2~P_n(\cos {\theta _2})~d{\theta _2}},
\end{align}
 where $u_2(\theta_2)$ determines the extent of the active section of the motor. At any arbitrary point ${\bf{r}}_0$ in free space, the concentration field generated by the first colloidal motor can be expressed up to the first term (dipole)
 \begin{eqnarray}
{n_2} = {\left. {{n_2}} \right|_{r = {r_0}}} + {\left. {{\bf{r}}_1 \cdot \nabla {n_2}} \right|_{r = {r_0}}} + ...,
 \end{eqnarray}
 where ${\bf{r}}_1$ is the position vector located at the center of mass of particle $1$. The gradient is in Cartesian coordinates which can be written as
 \begin{eqnarray}
 \nabla  = \frac{\partial }{{\partial x}}{{\bf{e}}_x} + \frac{\partial }{{\partial y}}{{\bf{e}}_y} + \frac{\partial }{{\partial z}}{{\bf{e}}_z},
 \end{eqnarray}
 where the partial derivatives can be transformed into spherical coordinates according to the following relations derived from the chain rule
 \begin{align}
&\frac{\partial }{{\partial x}} = \cos \phi \sin \theta \frac{\partial }{{\partial r}} + \frac{{\cos \phi \cos \theta }}{r}\frac{\partial }{{\partial \theta }} - \frac{{\sin \phi }}{{r\sin \theta }}\frac{\partial }{{\partial \phi }},\\
&\frac{\partial }{{\partial y}} = \sin \phi \sin \theta \frac{\partial }{{\partial r}} + \frac{{\sin \phi \cos \theta }}{r}\frac{\partial }{{\partial \theta }} + \frac{{\cos \phi }}{{r\sin \theta }}\frac{\partial }{{\partial \phi }},\\
&\frac{\partial }{{\partial z}} = \cos \theta \frac{\partial }{{\partial r}} - \frac{{\sin \theta }}{r}\frac{\partial }{{\partial \theta }}.
 \end{align}
For axisymmetric motion, there is no $\phi$ dependency. Also, since we are evaluating the derivatives at $\theta=0^{\circ}$ or $180^{\circ}$, all partial derivatives with respect to $\theta$ vanish and the concentration field created by colloid $2$ at the location of colloid $1$ is
 \begin{eqnarray}
{n_2} = \zeta_0 + \zeta_1{r_1}\cos {\theta _1},
 \end{eqnarray}
 where by evaluation we have,
 \begin{align}
&\zeta _0={\left. {{n_2}} \right|_{\theta  = 0,r = \Delta }} = \sum\limits_{n = 0}^\infty  {\frac{{{d_n}}}{{{\Delta ^{n + 1}}}}{P_n}(1)}, \\
&\zeta_1={\left. {\frac{{\partial {n_2}}}{{\partial r}}} \right|_{\theta  = 0,r = \Delta }} =  - \sum\limits_{n = 0}^\infty  {\frac{{(n + 1){d_n}}}{{{\Delta ^{n + 2}}}}{P_n}(1)}. 
\end{align}
 The concentration field around colloid $B$ by linear superposition can be established according to
 \begin{eqnarray}
 {n_1} = \sum\limits_{n = 0}^\infty  {\frac{{{f_n}}}{{r_1^{n + 1}}}{P_n}(\cos {\theta _1})} +\zeta_0 +\zeta_1{r_1}\cos {\theta _1},\label{nn1}
 \end{eqnarray}
 where $f_n$ can be determined by applying the boundary condition at the surface of colloid $B$ defined as
 \begin{eqnarray}
 {\left. {\frac{{\partial {n_1}}}{{\partial {r_1}}}} \right|_{{r_1} = 1}} =  - {\vartheta }{u_1(\theta_1)}.
 \end{eqnarray}
Thus the constant $f_n$ is found to be
 \begin{align}
&{f_n} = \frac{{2n + 1}}{{2(n + 1)}}\int_0^\pi  {{\vartheta }{u_1}\sin {\theta _B}~{P_n}(\cos {\theta _B})~d} {\theta _B},~~(n \ne 1)\\
 &{f_1} = \frac{3}{4}\int_0^\pi  {{\vartheta }{u_1}\sin {\theta _B}\cos {\theta _B}~d {\theta _B}} + \frac{{{\zeta _1}}}{2},
\end{align}
 where $\vartheta =\frac{N_1}{N_2}$.\\
 \indent Having evaluated the disturbed concentration field around colloid $1$, we can expand this solution around the center of mass of colloid $2$ similarly to have
 \begin{eqnarray}
 {n_1} = {\epsilon _0} - {\epsilon _1}{r_2}\cos {\theta _2},
 \end{eqnarray}
 where
 \begin{align}
&\epsilon _0={\left. {{n_1}} \right|_{\theta  = \pi,r = \Delta }} = \sum\limits_{n = 0}^\infty  {\frac{{{f_n}}}{{{\Delta ^{n + 1}}}}{P_n}(-1)}, \\
&\epsilon_1={\left. {\frac{{\partial {n_1}}}{{\partial r}}} \right|_{\theta  = \pi,r = \Delta }} =  - \sum\limits_{n = 0}^\infty  {\frac{{(n + 1){f_n}}}{{{\Delta ^{n + 2}}}}{P_n}(-1)}.
\end{align}
Consequently, the distorted concentration field around colloid $2$, by linear superposition, can be established:
 \begin{eqnarray}
 {n_2} = \sum\limits_{n = 0}^\infty  {\frac{{{g_n}}}{{r_2^{n + 1}}}{P_n}(\cos {\theta _2})} +\epsilon_0 -\epsilon_1{r_2}\cos {\theta _2},\label{nn2}
 \end{eqnarray}
  where $g_n$ can be determined by applying the boundary condition at the surface of colloid 2, which is defined as
 \begin{eqnarray}
 {\left. {\frac{{\partial {n_2}}}{{\partial {r_2}}}} \right|_{{r_2} = 1}} =  - {u_2}.
 \end{eqnarray}
Therefore, the constant $g_n$ is found to be
 \begin{align}
&{g_n} = \frac{{2n + 1}}{{2(n + 1)}}\int_0^\pi  {{u_2}\sin {\theta _2}~{P_n}(\cos {\theta _2})~d} {\theta _2},~~(n \ne 1)\\
 &{g_1} = \frac{3}{4}\int_0^\pi  {{u_2}\sin {\theta _2}\cos {\theta _2}~d {\theta _2} }- \frac{{{\epsilon _1}}}{2}.
\end{align}
Inasmuch as the flow is axisymmetirc, the colloids have no angular velocity. Here, we only focus on the single body motion and ignore pair hydrodynamic interactions and therefore for the rest of the text we drop the subscript $1$ and $2$ for the coordinates. To do so, we suppose we have a colloidal motor with radius $a$ which diffusiophoretically translates along the $z$-axis due to a concentration gradient described in previous sections. We then decompose this problem into two problems: (I) the translational motion of the colloid with constant velocity in the $z$-direction ${\bf{U}}=U {\bf{e}}_z$ and (II) the colloid is stationary but the fluid slips on the particle surface. The total force $\bf{F}$ on the colloid can be found by superposition of the force obtained from the two problems and it should be equal to zero,
\begin{eqnarray}
{{\bf{F}}_T} = {{\bf{F}}_1} + {{\bf{F}}_2} = 0.\label{FT}
\end{eqnarray}
The first problem is the classical Stokes flow problem and can be solved by stream function expansion which ultimately yields the force on the particle:
\begin{eqnarray}
{{\bf{F}}_1} = -6\pi \mu Ua~{{\bf{e}}_z}.\label{F1}
\end{eqnarray}
 For the second problem, we use RRT (Eq.~\ref{rrt}) to calculate the force. In order to do this, we let ($\bf{v}''$, $\sigma''$) be the solution of the classical Stokes flow problem which satisfies the following boundary condition on particle surface:
  \begin{eqnarray}
 {\bf{v''}}(r = a) = {{\bf{e}}_z}.
 \end{eqnarray}
The corresponding stress tensor $\sigma''$ is
\begin{eqnarray}
{{\sigma ''}_{rr}} =  - {p_\infty } + \frac{3}{2}\mu \cos \theta \left(\frac{a^2}{{{}{r^3}}} + \frac{a}{{2{r^2}}} - \frac{a^3}{{2{r^4}}}\right),
\end{eqnarray}
\begin{eqnarray}
{{\sigma ''}_{r\theta }} = {{\sigma ''}_{\theta r}} =  - \frac{3}{2}\frac{\mu a^3}{{{r^4}}}\sin \theta,
\end{eqnarray}
where $p_{\infty}$ is a constant pressure far away from the colloid and, in this case, Eq.~ \ref{rrt} can be simplified to
\begin{eqnarray}
F_2^z = \mathop{{\int\!\!\!\!\!\int}\mkern-21mu \bigcirc}\limits_{\partial \Sigma } 
 {{{\bf{e}}_r} \cdot \sigma '' \cdot {{\bf{v}}_s}~dS}.
\end{eqnarray}
 In spherical coordinates, we can write the slip velocity as,
 \begin{eqnarray}
 {\nabla _s } = \frac{1}{r}\frac{\partial }{{\partial \theta }}{\bf{e}_{\theta}} + \frac{1}{{r\sin \theta }}\frac{\partial }{{\partial \phi }}{\bf{e}_{\phi}} ,
 \end{eqnarray}
 and therefore we have
\begin{eqnarray}
{{\bf{v}}_s} =  - \frac{{bn_c}}{r}\frac{{\partial n}}{{\partial \theta }}{\bf{e}}_{\theta},
\end{eqnarray}
\begin{eqnarray}
\frac{{\partial n}}{{\partial \theta }} = \sum\limits_{n = 0}^\infty  {\frac{{{a^{n + 1}}{A_n}}}{{{r^{n + 1}}}}\frac{{d{P_n}(\cos \theta )}}{{d\theta }}}  \pm \frac{{{\lambda _1}r}}{a}\sin \theta,
\end{eqnarray}
where $n$ can be governed by Eqs.~\ref{nn1} and \ref{nn2} and therefore $A_n$ and $\lambda$ represent either $f_n$ or $g_n$ and  $\zeta_1$ or $\epsilon_1$, respectively. Furthermore, we have,
\begin{align}
&{{\bf{e}}_r} \cdot \sigma '' \cdot {{\bf{v}}_s} =\nonumber\\ &-{{\sigma ''}_{r\theta }}\frac{{b{{n}_c}}}{r}\frac{{\partial n}}{{\partial \theta }} = \frac{3}{2}\frac{{{a^3}\mu }}{{{r^4}}}\sin \theta \frac{{b{{n}_c }}}{r}\left[\sum\limits_{n = 0}^\infty  {{A_n}\frac{{{a^{n + 1}}}}{{{r^{n + 1}}}}\frac{{d{P_n}(\cos \theta )}}{{d\theta }}}\pm \frac{{{\lambda _1}r}}{a}\sin \theta \right],
\end{align}
\begin{eqnarray}
F_2^z = \mathop{{\int\!\!\!\!\!\int}\mkern-21mu \bigcirc}\limits_{\partial \Sigma } 
 {{{\bf{e}}_r} \cdot \sigma '' \cdot {{\bf{v}}_s}dS}  = 3\pi \mu b{n_c}\int_0^\pi  {\left[\sum\limits_{n = 0}^\infty  {{A_n}{{\sin }^2}\theta \frac{{d{P_n}(\cos \theta )}}{{d\theta }}}  \pm {\lambda _1}{{\sin }^3}\theta \right]d\theta },
\end{eqnarray}
\begin{eqnarray}
{\sin ^2}\theta \frac{{d{P_n}(\cos \theta )}}{{d\theta }}d\theta  = (1 - {\eta ^2})\frac{{d{P_n}(\eta )}}{{d\eta }}d\eta  = n({P_{n - 1}}(\eta ) - \eta {P_n}(\eta ))d\eta ,,
\end{eqnarray}
where we introduce
\begin{eqnarray}
\eta=\cos \theta.
\end{eqnarray}
One can find that
\begin{eqnarray}
F_2^z = 3\pi \mu b{{n}_c }\int_{ - 1}^1 {\sum\limits_{n = 0}^\infty  {n{A_n}({P_{n - 1}}(\eta ) - \eta {P_n}(\eta ))d\eta } }  =  4\pi \mu b{{n}_c}[{A_1}\pm \lambda_1].
\end{eqnarray}
Finally by utilizing Eqs.~\ref{FT} and \ref{F1} we find that the diffusiophoretic velocity is given by
\begin{eqnarray}
U = \frac{{2b{n_c}({A_1} \pm \lambda_1 )}}{{3a}}.
\end{eqnarray}
the above analysis would be highly simplified for the cases of two catalytically active colloids and a catalytic colloid and a cargo (inert colloid). 
\subsection{Two catalytic colloids}
Suppose we have two catalytic  colloids with radius $a$ and, due to a chemical reaction, they both generate a surface flux of $N_1$ and $N_2$, respectively. The non-dimensional concentration field around the first colloid in the absence of the second colloid is
\begin{eqnarray}
{n_2} = \frac{1}{r_2}.
\end{eqnarray}
By expanding this solution at the second colloid's center of mass, one can show that
\begin{eqnarray}
{n_2} = \frac{1}{\Delta } - \frac{{{r_1}\cos {\theta _1}}}{{{\Delta ^2}}}.
\end{eqnarray}
The concentration field in this case around the second colloid is
\begin{eqnarray}
{n_1} = \frac{{{\vartheta }}}{r_1} + \frac{1}{\Delta } - \frac{{\cos {\theta _1}}}{{{\Delta ^2}}}\left[r_1+ \frac{1}{{2{r_1^2}}}\right].
\end{eqnarray}
A similar procedure for colloid $1$ yields a concentration field around colloid $2$
\begin{eqnarray}
{n_2} = \frac{1}{r_2} + \frac{{{\vartheta }}}{\Delta } + \frac{{{\vartheta }\cos {\theta _2}}}{{{\Delta ^2}}}\left[r_2 + \frac{1}{{2{r_2^2}}}\right].
\end{eqnarray}
Having utilized RRT, we find that the swimming velocities of the colloids are,
\begin{eqnarray}
{U_1} = -\frac{{b{n_c}{\vartheta}}}{{{r_{c1}}{\Delta ^2}}},\\
{U_2} =  \frac{{b{n_c}}}{{{r_{c2}}{\Delta ^2}}}.
\end{eqnarray}
\subsection{A catalytically active colloid and a cargo}
In this case, we let colloid $2$ be inert and impenetrable to the solute while the first colloid is catalytically active. Utilizing a similar procedure to the one above, the concentration field around the second colloid due to the presence of colloid $1$ is given by
\begin{eqnarray}
{n_2} =  \frac{1}{\Delta } - \frac{{\cos {\theta _2}}}{{{\Delta ^2}}}\left[r_2 + \frac{1}{{2{r_2^2}}}\right],
\end{eqnarray}
and the distorted concentration field around colloid $1$ due to the presence of colloid $2$ is
\begin{eqnarray}
{n_1} =   \frac{1}{{2{\Delta ^4}}} + \frac{1}{r} + \frac{\cos {\theta _1}}{\Delta^5}\left[{r_1} + \frac{1}{{2r_1^2}}\right].
\end{eqnarray} 
Taking advantage of the RRT, we determine that the colloid velocities are
\begin{align}
&{U_1} = \frac{{b{n_c}{\vartheta _2}}}{{{a}{\Delta ^5}}},\\
&{U_2} =  - \frac{{b{n_c}}}{{{a}{\Delta ^2}}}.
\end{align}
\newpage
% \bibliography{Ref}
\end{document}